\documentclass [manuscript]{aastex}
\usepackage{amsmath}
\usepackage{graphicx}
\usepackage{longtable}

\shorttitle{the absorption of stellar Ly $\alpha$ by the exoplanet's atmosphere}
\shortauthors{Yan \& Guo}

\begin{document}

\title{THE ESCAPE OF HYDROGEN-RICH ATMOSPHERE OF EXOPLANET: MASS LOSS RATES AND THE ABSORPTIONS OF STELLAR LYMAN $\alpha$ }


\author{D. D. Yan\altaffilmark{1,2,3} }
\affil{Yunnan Observatories,
Chinese Academy of Sciences, P.O. Box 110, Kunming 650011, China}

\author{J. H. Guo\altaffilmark{1,2,3} }
\affil{Yunnan Observatories,
Chinese Academy of Sciences, P.O. Box 110, Kunming 650011, China; guojh@ynao.ac.cn}

\altaffiltext{1}{Yunnan Observatories,
Chinese Academy of Sciences, P.O. Box 110, Kunming 650011, China}
\altaffiltext{2}{School of Astronomy and Space Science, University of Chinese Academy of Sciences, Beijing, China}
\altaffiltext{3}{Key Laboratory for the Structure and Evolution of
Celestial Objects, CAS, Kunming 650011, China}

\begin{abstract}
Since the mass loss rates are the function of the mean density of the planet and the stellar irradiation, we calculated about 450 models covering planets with different densities and stellar irradiation. Our results show that the mass loss rates are dependent on the stellar irradiation and the mean density. However, the mass loss rates predicted by the energy-limited equation are higher than that of hydrodynamic model when the XUV integrated flux is higher than $\sim$2$\times$10$^{4}$ erg/cm$^{2}$/s. The overestimation can be revised if the kinetic and thermal energy of the escaping atmosphere is included in the energy-limited equation. We found that the heating efficiencies are proportional to the product of the gravitational potential of the planet and the stellar irradiation. The mean absorption radii of stellar irradiation are 1.1-1.2 R$_{p}$ for the Jupiter-like planets while they vary in the range of 1.1-1.7 R$_{p}$ for the planets with smaller sizes. We evaluated the absorption of stellar Ly$\alpha$ by planetary atmosphere and found that the deeper Ly$\alpha$ absorptions tend to locate in the high stellar irradiation and low planetary mean density regions, and vice versa. Moreover, planets with mass loss rates higher than 10$^{11} g/s$ are likely to exhibit obvious absorptions. Finally, we suggested that the absorption levels are related to the inherent properties of the exoplanets. The planets with larger sizes (or lower mean density) show strong Ly$\alpha$ absorptions. Neptune-like and Earth-like planets tend to have weak Ly$\alpha$ absorptions because of their small sizes (or high densities).

\end{abstract}


\section{Introduction} \label{sec:intro}

Exoplanets orbiting close to their host stars are subjected to the intense X-Ray (1$\sim$100$\rm\AA$) and Extreme Ultraviolet (100$\sim$912$\rm\AA$) radiation (hereafter XUV) from their host stars. Due to the intense XUV radiation, gas-rich planets are likely to suffer from hydrodynamic atmospheric escape that would push the gas beyond their Roche lobes \citep{2003APJL...598..L121,2007Natur...450..854,2013Icar...226..1678}. By studying the process of the escape of atmosphere, we can understand the composition and structure of exoplanetary atmosphere. Understanding such process is also imperative to probe the habitability of exoplanets because water can be lost through hydrodynamic escape \citep{1996JGR...101..26039,2007Icar...191..453,2010MNRAS...409..963,2013APJ...765..131,2017arxiv...1705..05535, 2019ApJ...872..99}.

Observations among transit systems have revealed excess stellar spectrum absorption apart from the optical occultation by planet itself. The detection in ultraviolet Ly$\alpha$ of hot-Jupiter HD 209458b was the first transit case in which \citet{2003Nature...422..143} found a $\sim$15\% Ly$\alpha$ absorption by analysing the data of HST/STIS observations with medium-resolution. The absorption of Ly$\alpha$ was  confirmed by \citet{2007ApJ...671..L61,2008ApJ...688..1352}, who demonstrated a lower absorption $\sim$8.9$\%$ $\pm$ 2.1$\%$. Subsequently, \citet{2010ApJ...709..1284} reanalysed the HST/STIS observations of HD 209458b and proposed the Ly$\alpha$ absorption depth in wavelength range [1212 $\rm\AA$, 1220 $\rm\AA$] was 6.6$\%$ $\pm$ 2.3$\%$. However, the optical occultation by HD 209458b is only $\sim$1.5\% \citep{2000APJ...529..L41,2000APJ...529..L45}. Thus the expanded natural hydrogen around the planet is a good interpretation for the excess absorption. The excess absorption of Ly$\alpha$ has also been detected in HD 189733b and GJ 436b. For HD 189733b, an absorption of $\sim$5$\%$ has been detected by \citet{2010A&A...514..72,2012A&A...543..L4} and \citet{2013A&A...551A..A63}. Observations of warm-Neptune GJ 436b also indicate that it has an expanded exosphere  \citep{2014APJ...786..132}. It was detected that a large comet-like tail of hydrogen was surrounding GJ 436b, and this planet obscured almost 50$\%$ about the Ly$\alpha$ of its host star \citep{2011A&A...529..A80,2015Nature...522..459}.

Moreover, \citet{2007Nature...445..511} revealed an excess $\sim$0.03\% absorption in the Balmer jump and continuum from HD 209458b. \citet{2012ApJ...751..86} reported the detection of H$\alpha$ excess absorption for HD 189733, which hints that there are the excited hydrogen atoms in its atmosphere. Recently, \citet{2018NatAs...2..714Y} found the excess absorption of H$\alpha$ in KELT-9b. Due to the very high temperature of KELT-9b (4600K), the excited atoms can produce an extended hydrogen envelope of 1.64 planetary radius, which implies the escape of hydrogen. The excess absorptions are also found in helium and heavier elements. \citet{2004APJ...604..L69} detected O $_{\rm\uppercase\expandafter{\romannumeral1}}$, C $_{\rm\uppercase\expandafter{\romannumeral2}}$ in the atmosphere of HD 209458b. Subsequently, \citet{2013A&A...553A..A52} also detected oxygen atoms and possibly C $_{\rm\uppercase\expandafter{\romannumeral2}}$ in the upper atmosphere.

Generally, the excess absorptions are attributed to the escape of gas in the extended envelope. Hydrodynamical simulations including the process of radiative transfer and photochemistry are implemented to study the mechanism of atmospheric escape \citep{2004Icar...170..167,2005APJ...621..1409,2007P&SS...55..1426,2008P&SS...56..1260,2009APJ...693..23,2011ApJ...733..98, 2013ApJ...766..102,
2016MNRAS...460..1300, 2016A&A...586..75, 2018A&A...619A.151K, 2018ApJ...866...47S}, and the observational absorption of Ly$\alpha$ for HD 209458b and HD 189733b can be explained at some extent by such model \citep{2013Icar...226..1678,2016ApJ...818..107,2019arXiv190310772O}. On the other hand, it is believed that planetary magnetic field would play an important role in controlling the atmospheric escape \citep{2010APJ...709..670,2011APJ...730..27,2011APJ...728..152,2012APJL...753..L4,2014MNRAS...444..3761,2015APJ...813..50,2017MNRAS.470.4330E,2019MNRAS.483.2600D}.

Thousands of exoplanets have been discovered to date, however, only a few among them have been detected undergoing atmospheric escape and most  theoretical studies focus on the well-known observed transit systems as mentioned above. It is crucial to find more information about the latent and unexplored atmospheric escape. To explore the properties of the escaping atmosphere, it is necessary  to know the dependence of the mass loss rates on the physical parameters of exoplanets since the mass loss rates can describe the levels of the absorption of Ly$\alpha$ in a certain extent (The absorptions also depend on the degree of ionization). The mass loss rates of exoplanet are related to many physical parameters, such as the masses and radii etc. However, at a large extent the mass loss rates are determined by the XUV fluxes received by the planets and the mean densities of the planets, namely, $\dot{M}\varpropto  \frac{F_{xuv}}{\rho}$ \citep{2003APJL...598..L121,2009AA...506..399}. The equation of energy-limit presented by \citep{2003APJL...598..L121,2009AA...506..399} have been generally used to estimate the mass loss rates. This hints that one can obtain the general trend of mass loss rates if the distributions of XUV fluxes and the mean densities of many planets are known.

In this paper, we aim to compare the properties of atmosphere of exoplanet with different XUV flux and density and inspect whether the energy-limited equation is suitable for them. Furthermore, we investigate the absorption of stellar Ly$\alpha$ by the exoplanets' atmosphere for a variety of samples ranging from Earth-Like planets to Jupiter-like planets. As an exploring work, the absorption by interstellar medium is not included because a goal of this paper is to discuss how the stellar Ly$\alpha$ is absorbed by the atmosphere of exoplanet rather than predicting the observable signals. To discuss the properties of atmosphere and the absorption of Ly$\alpha$ by the planetary atmosphere, we select some sample from those planets that have been confirmed (Section. 2.1) and calculate the mass loss rates in a large parameter space. We obtain the XUV spectra by using the method of \citet{2011A&A...532..6}(Section 2.2). The hydrodynamic model and the calculation of Ly$\alpha$ absorption are presented in Section 2.3 and 2.4. In Section \ref{sec:result2}, we present the results of our selected sample and give a statistical analysis of them. In Section \ref{sec:discu}, we discuss the limitations of our work. Finally in Section \ref{sec:summ}, we summarize the results.

\section{Method and  Model} \label{sec:memo}

\subsection{Sample selected } \label{subsec:samples}

The observations show that the radii of most exoplanets are smaller 2R$_J$ (R$_J$ is the radius of Jupiter.) (http://exoplanet.eu). Thus, we confined the planets in the range of the radius less than 2 R$_J$. In addition, for exoplanets with high mass their atmospheres can be compact and the escape of species is relatively difficult. We thus set the upper limit of the mass to 2 M$_J$ (M$_J$ is the mass of Jupiter). Finally, we further selected the sample by their gravitational potentials. The calculations of \citet{2016A&A...585..2s} found that the hydrodynamic escape of the atmosphere is difficult for the exoplanets with high gravitational potentials ($>$4$\times$10$^{13}$ erg g$^{-1}$), which means that in order to produce the hydrodynamic escape the exoplanets with high mass should have large radius (or relatively low mean densities). Thus, the gravitational potential of the sample planets are smaller than 4$\times$10$^{13}$ erg g$^{-1}$.

According to \citet{2014ApJ...794..133}, planets with radii 0.0885-0.177 R$_J$ (R$_J$  is the radius of Jupiter) are Super-Earths, 0.177-0.354 R$_J$ are Mini-Neptunes and 0.354-0.531 R$_J$ are Super-Neptunes. In the investigation, we also classified the planets with different sizes. Specifically, the planets with radius smaller than 0.2 R$_J$ are Earth-like planets, 0.2 -0.6 R$_J$ are Neptune-like planets, 0.6-1.0 R$_J$ are Saturn-like planets. Finally, the planets with radius larger than 1.0 R$_J$ are Jupiter-like planets.

\citet{2007Natur...450..854} suggests that the exoplanets with a separation smaller than 0.15AU will produce significant mass loss. Hence, we selected those planets that their separation are less than 0.1 AU. The separations are in the range of 0.01-0.09AU for Earth-like and Neptune-like planets,  0.04-0.07 AU for Saturn-like planets and 0.02-0.08AU for Jupiter-like planets. The host stellar masses of each system are in the range of 0.08-1.105 M$_{\sun}$ (M$_{\sun}$ is the mass of the Sun) for Earth-like, 0.15-1.223 M$_{\sun}$ for Neptune-like, 0.816-1.3 M$_{\sun}$ for Saturn-like and 0.8-1.46 M$_{\sun}$ for Jupiter-like planets. Even though the stellar mass in some Earth-like and Neptune-like systems are relatively small, these systems account for a small proportion in each group.

Based on the conditions above, we selected 90 exoplanets from the real systems (http://exoplanet.eu and https://exoplanetarchive.ipac.caltech.edu/), and some artificially made planets will be added by the real planets. For the sake of a diversity, we investigated planets from Earth-sized to Jupiter-sized. We emphasize that it is not the goal of this paper to predict real mass loss rates and the Ly$\alpha$ absorptions of real planets because the accurate characterizations are difficult due to the lack of some important physical input of the host stars. For example, the XUV spectral energy distributions (SEDs) can not be obtained easily because they can not be well determined from observations due to the obscure of ISM. In addition, the compositions of Earth-like planets can be different from the hydrogen-dominated atmosphere of giant planets. Here we assumed that they hold a hydrogen-rich envelope. The hydrogen of their atmospheres can originate from the dissociation of H$_{2}$O \citep[]{1983Icarus...53..479,2019ApJ...872..99}. Furthermore, the hydrogen-rich atmosphere can appear in the early phase of Earth-like planets due to the process of outgassing or accretion \citep[]{2008APJ...685..1237}. The observational signals in Ly$\alpha$ for Earth-like planets have been discussed by \citet{2019A&A...622..46} and \citet{2019A&A...623A.131K}.


%

\begin{figure}
\begin{minipage}[t]{0.5\linewidth}
\centering
\includegraphics[width=3.6in,height=3.in]{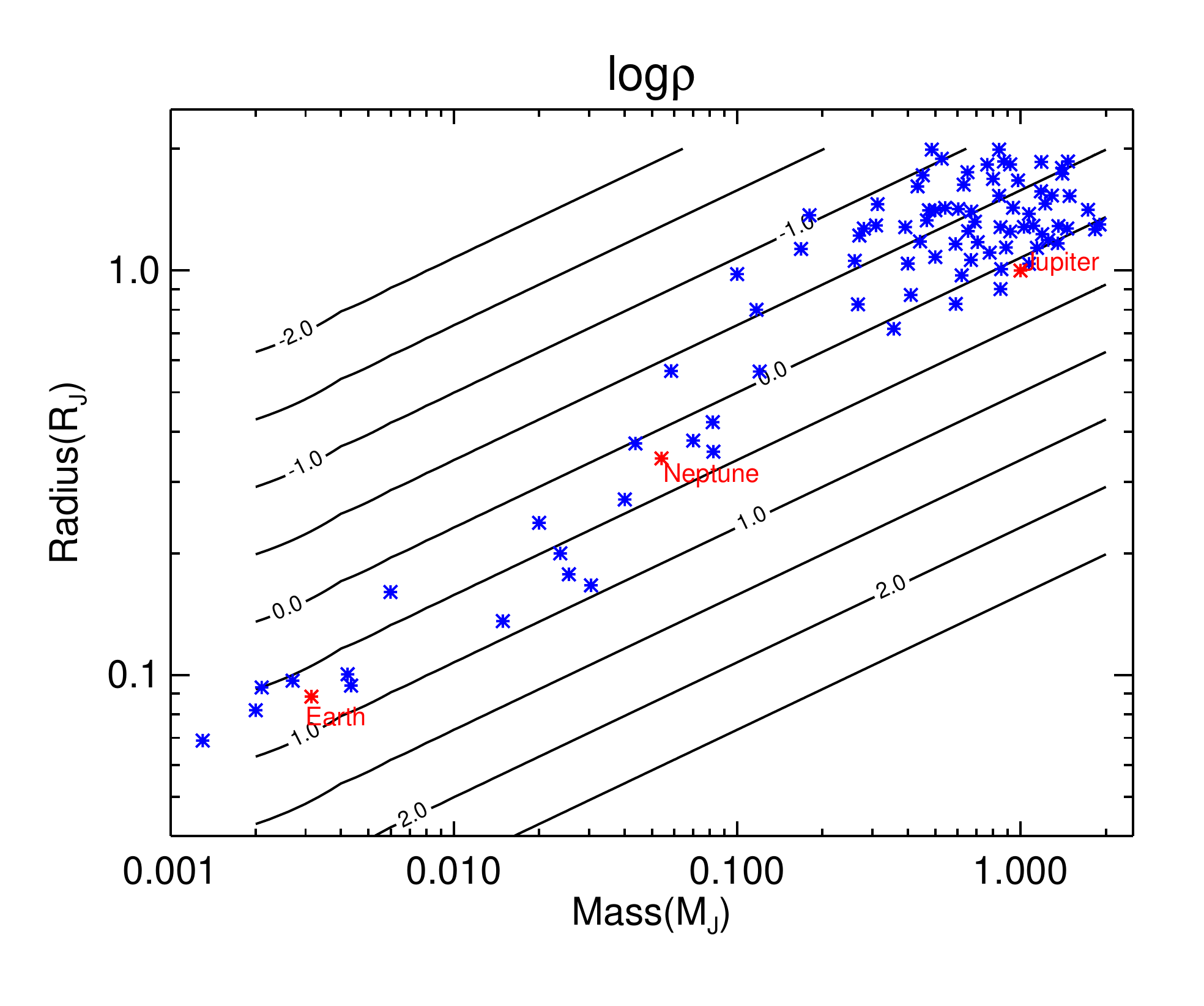}
\end{minipage}
\begin{minipage}[t]{0.5\linewidth}
\centering
\includegraphics[width=3.6in,height=2.9in]{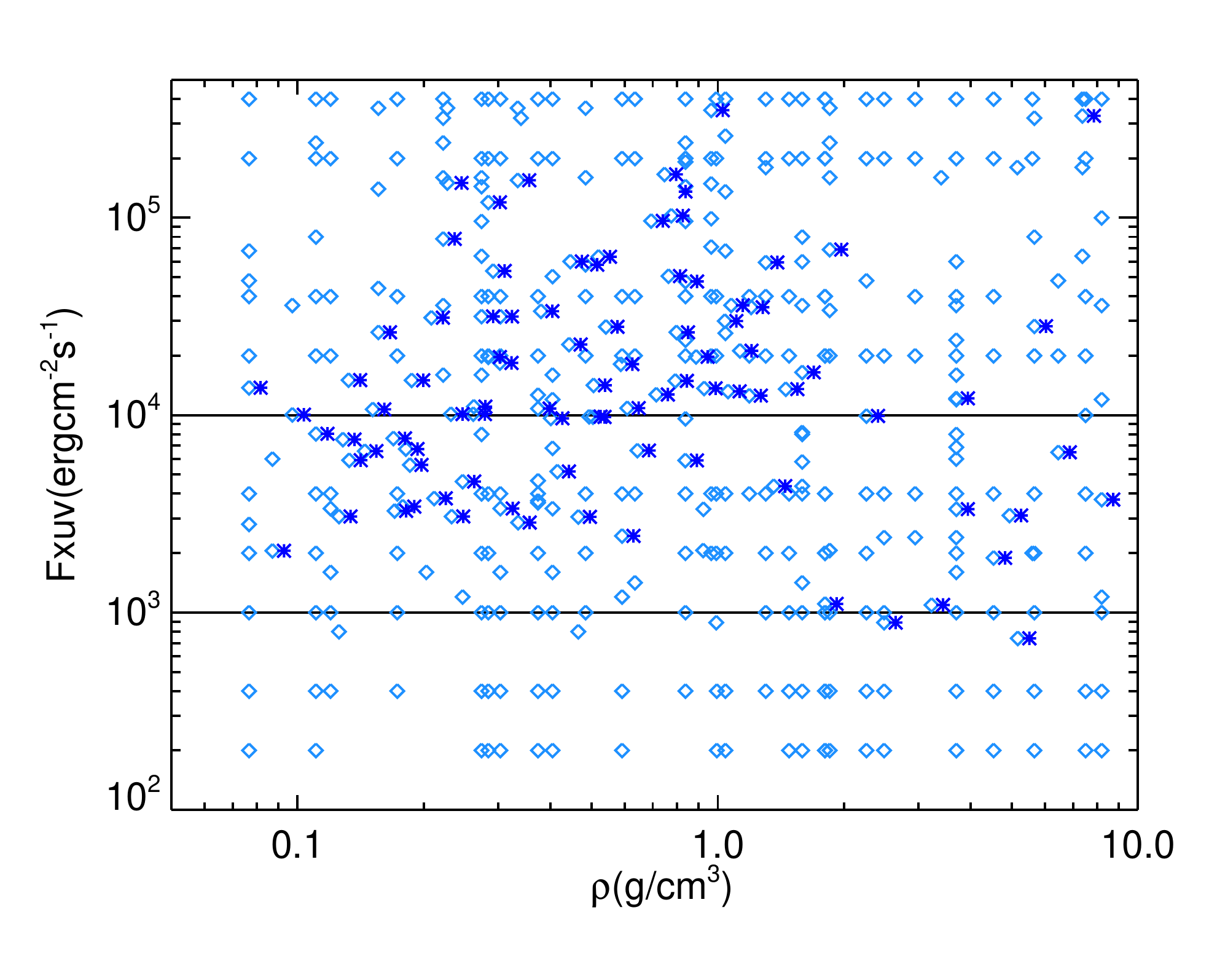}
\end{minipage}
\caption{The sample of this work. In the left panel, the x-axis is the planetary mass, the y-axis is the planetary radius and the contours are the planetary mean density plotted in log$_{10}$ form. The planets in the left panel are the observed ones. In the right panel, the x-axis is the planetary mean density and y-axis is integrated XUV flux. The dark blue asterisks represent the planets corresponding to the same planets in the left panel and the light blue diamonds represent the artificially made planets.\label{sample12} }
\end{figure}

The left panel of Figure. 1 shows the mass and radius of the sample planets (In the panel log $\rho$ expresses the logarithm  of the planetary mean density). One can see that our sample covers various exoplanets ranging from Jupiter-like planets to Earth-like planets. We  neglected the Earth-like planets with large radius because such planets may be composed of gas rather than rock. As discovered by Lammer et al. (2003, 2009), the mass loss rates are controlled essentially by the XUV flux and the mean density of the planets. Furthermore, the Ly$\alpha$ absorption by the planetary atmosphere can be estimated after the mass loss rates are obtained. Therefore, we plot the sample in the Fxuv-$\overline{\rho}$  diagram (The right panel in Figure 1. For more details of Fxuv, see Section \ref{subsec:xuv}). The dark blue asterisks represent the planets of the Radius-Mass diagram. As we can see, most planets locate at the center of the Fxuv-$\overline{\rho}$ diagram and their mean densities are similar or lower than that of Jupiter. The planets with high density locate at the right of the panel. In order to investigate the the influence of Fxuv and $\overline{\rho}$ on the mass loss rates and absorption of stellar Ly$\alpha$ by planetary atmosphere on a larger Fxuv-$\rho$ scale, we added many artificially made planets which are represented by the light blue diamonds. Most of the artificial ones are made by changing the input Fxuv of the planets and the rest of those planets is made by changing the planetary mass and radius. Most of mock planets made by changing planetary masses and radii are Earth-like planets or Neptune-like planets, as the initial number of planet selected in this range is less than Jupiter-like planets. By modifying the masses and radii of those Earth-like to Neptune-like planets, the sample planets cover uniformly the Fxuv-$\overline{\rho}$  diagram. In total, about 450 planets are included.

In our solar system, the mean density of the rocky planets are higher than those of the gaseous planets. The distributions of the mean densities of our sample are similar with that of the solar system. In our sample the mean densities of Jupiter-like and Saturn-like planets are in the range of 0.08-1.4 g/cm$^{3}$. For the Neptune-like planets, the mean densities are in the range of 0.3-2.65 g/cm$^{3}$. The smallest density of Earth-like planet is 0.9 g/cm$^{3}$ while the largest density is 8.2 g/cm$^{3}$. A such distribution means that the densities of most Jupiter-like planets are lower than those of Earth-like planets. Furthermore, the distributions of the gravitational potentials are different for different planets. For earth-like and Neptune-like planets, their gravitational potentials are smaller than 3$\times$10$^{12}$ erg g$^{-1}$. However, the gravitational potentials of Saturn-like and Jupiter-like planets cover a large range (1$\times$10$^{12}$-3$\times$10$^{13}$ erg g$^{-1}$) and are generally greater than those of planets with smaller sizes although there is overlap around 3$\times$10$^{12}$ erg g$^{-1}$. The highest flux we set is 4$\times$ 10$^{5}$ erg/cm$^{2}$/s, which corresponds a planet orbiting the Sun at 0.003 AU. The lowest value of the XUV flux is 2$\times$ 10$^{2}$ erg/cm$^{2}$/s below which the occurrences of hydrodynamic escape can be difficult. Our calculations cover a large range in the Fxuv-$\rho$ diagram. One can see from Fxuv-$\rho$ diagram that in any given level of fluxes there are many planets with different mean density. By investigating the dependence of the mass loss rates on the XUV flux and means density, we can further conclude the general trend of Ly$\alpha$ absorption and explore what factors determine the absorption levels.

\subsection{The XUV spectra of stars } \label{subsec:xuv}
As discussed above, the mass loss rates are related with the properties of planet and the XUV irradiation. It is difficult to obtain the accurate spectra of star because ISM can obscure the EUV emission of the stars. \citet{2011A&A...532..6} fitted the observation for late F to early M dwarfs and found that the XUV luminosity (integrated XUV flux received at the planetary orbits) can be depicted by an empirical relation:
\begin{equation}
logL_{euv}=(29.12\pm0.11)-(1.24\pm0.15)logt
\end{equation}

\begin{equation}
L_{x}=\left\{
     \begin{array}{lcl}
      6.3\times10^{-4}L_{bol} (t \leq t_i)  \\
      1.89\times10^{28}t^{-1.55}(t \textgreater t _i)
      \end{array}
      \right.
\end{equation}
with t$_i$=2.03$\times$10$^{20}$L$_{bol}$ $^{-0.65}$, where t is the stellar age in Gyr, L$_{euv}$ and L$_{x}$ (erg/s) are the luminosity in EUV band and Xray band, respectively.

This means that the X-ray and EUV integrated fluxes can be obtained if the age of the star is known. In this paper, we only choose the systems with given ages and the ages are obtained from http://exoplanet.eu. Even though the ages usually given with relatively large uncertainties, an accurate age is not necessary because the purpose of the present work is to explore the response of the mass loss rates and the Ly$\alpha$ absorption depth to XUV flux. Thus, the age only provide a XUV flux estimation. In order to simulate the emission the XUV spectra of stellar corona, we used the software of XSPEC-APEC in which the wavelength domain of 1 $\rm\AA$ to 912 $\rm\AA$ was used. The free parameters in APEC are metallicity and coronal temperature. We first set the metallicity to solar value \citep{2009ARA&A...47..481}, so the spectra only depend on the coronal temperature. To obtain the XUV spectra of all stars, we calculated lots of spectra by using XSPEC-APEC. By comparing the theoretic and empirical ratio of Lx to Leuv in a given age, the optimum coronal temperatures of the stars can be defined. Finally, the XUV spectra of all stars are obtained by the method. We inspected the profiles of those XUV spectra and found that the $\beta$ indexes of all spectra are $\sim$ 0.9 ($\beta$=F$_{1-400{\rm\AA}}$/F$_{1-912{\rm\AA}}$, for details see \citet{2016ApJ...818..107} ). Thus, the influence of the profiles of the XUV spectra is slight (We evaluated the variations of the mass loss rates produced by the different profiles and found the change is smaller than 5\%.). We emphasized that the XUV fluxes obtained by the Equation (1) and (2) can result in deviation for those real planets due to the uncertainties of age and the models. However, the empirical spectra can express the essential feature of those stars. Thus, as a theoretical exploration the spectra can be used to describe the response of the atmosphere of planet on the different levels of Fxuv. In the premise, we will explore how the mass loss rates and the absorptions of Ly$\alpha$ are affected by the XUV flux of the stars and the properties of the planets.

\subsection{Hydrodynamic simulations } \label{subsec:hdsim}
We used the 1D  hydrodynamic models \citep{2018CHA&A...42..81} to simulate the atmospheric structures of our selected planet sample. Compared to the early models of \citet{2013ApJ...766..102} and \citet{2016ApJ...818..107}, there are two improvements in the model of \citet{2018CHA&A...42..81}. First, the former models use the solar EUV (100-912 $\rm\AA$) as stellar radiation spectra, while in the later model developed by \citet{2018CHA&A...42..81}, the radiation spectra expand to XUV (1-912 $\rm\AA$). The photoionization cross sections in X-Ray are smaller than those of EUV, so the heating by X-Ray is not remarkable compared with EUV. However, the photoionization produced by X-ray can be important for heavy elements. Second, the former ones only include hydrogen, helium and electrons, but the later one also includes heavier elements such as C, N, O, Si. To be specific, the later model includes 18 kinds of particles, among which are 7 kinds of  neutral particles, 10 kind of ions and electrons. In such condition, the photochemical reactions included photoionization, photodissociation, impact ionization, recombination, charge exchange and other important reactions (Table. 1).  For more details of the hydrodynamic model, the reader can refer to the papers above.

In the simulations, the metallicities of planets are consistent with those of their host stars. For those planets with unknown metallicities, we used the solar metallicity. In order to express the average of the energy received by the planet over the whole surface, the XUV fluxes are divided by a factor of 4. We choose the outer boundaries to be the host stars' radii for our sample. For the systems whose host stars' radii far exceed the planetary radii (such as Earth-like planet), the calculating outer boundaries are set to 15 times the planetary radii. In order to cover the surface of the host stars, the results are extrapolated to the radii of their host stars.

\begin{longtable}{lll}
\caption{The coefficients of the chemical reactions.}
\label{table:label} \\
 \hline
H${_2}$ + $h\nu$ $\rightarrow$ H${_2^+}$ + e  &    & \citet{2016ApJ...818..107}  \\
H${_2}$ + $h\nu$ $\rightarrow$ H${^+}$ + H + e  &   & \citet{2016ApJ...818..107} \\                                                                                                                                H + $h\nu$ $\rightarrow$ H${^+}$ + e  &    & \citet{2002APJ...575..33}   \\
He + $h\nu$ $\rightarrow$ He${^+}$ + e  &    & \citet{2002APJ...575..33}  \\
C + $h\nu$ $\rightarrow$ C${^+}$ + e  &   & \citet{1996ApJ...465..487V}\\
N + $h\nu$ $\rightarrow$ N${^+}$ + e  &    & \citet{1996ApJ...465..487V}   \\
O + $h\nu$ $\rightarrow$ O${^+}$ + e  &    & \citet{1996ApJ...465..487V}  \\
Si + $h\nu$ $\rightarrow$ Si${^+}$ + e  &   & \citet{1996ApJ...465..487V}   \\
Si$^+$ + $h\nu$ $\rightarrow$ Si$^{2+}$ + e  &   & \citet{1996ApJ...465..487V}  \\
H${_2}$ + M $\rightarrow$ H + H + M &   1.5 $\times$ 10$^{-9}$e$^{-48000/T}$ & \citet{1992JPCRD..21..411B} \\
H + H + M $\rightarrow$ H${_2}$ + M &   8.0 $\times$ 10$^{-33}$(300/T)$^{0.6}$ & \citet{1970JChPh..53.4395H}\\
H${_2^+}$ + H${_2}$ $\rightarrow$ H${_3^+}$ + H &   2.0 $\times$ 10$^{-9}$ & \citet{1974JChPh..60.2840T}\\
H${_3^+}$ + H $\rightarrow$ H${_2^+}$ + H${_2}$ &   2.0 $\times$ 10$^{-9}$ & \citet{2004Icar...170..167}\\
H${_2^+}$ + H $\rightarrow$ H${^+}$ + H${_2}$ &   6.4 $\times$ 10$^{-10}$ &  \citet{1979JChPh..70.2877K}\\
H${^+}$ + H${_2}$(v $\geq$ 4) $\rightarrow$ H${_2^+}$ + H &   1.0 $\times$ 10$^{-9}$e$^{-21900/T}$ & \citet{2004Icar...170..167}\\
He${^+}$ + H${_2}$ $\rightarrow$ HeH${^+}$ + H &    4.2 $\times$ 10$^{-13}$ & \citet{1989JChPh..91.4593S} \\
He${^+}$ + H${_2}$ $\rightarrow$ H${^+}$ + H + He &    8.8 $\times$ 10$^{-14}$ & \citet{1989JChPh..91.4593S}\\
HeH${^+}$ + H${_2}$ $\rightarrow$ H${_3^+}$ + He &    1.5 $\times$ 10$^{-9}$ & \citet{1980JChPh..73.4976B}\\
HeH${^+}$ + H $\rightarrow$ H${_2^+}$ + He &    9.1 $\times$ 10$^{-10}$ &  \citet{1979JChPh..70.2877K}\\
H${^+}$ + e $\rightarrow$ H + $h\nu$  &   4.0 $\times$ 10$^{-12}$(300/T)$^{0.64}$ & \citet{1995MNRAS.272...41S}\\
He${^+}$ + e $\rightarrow$ He + $h\nu$  &   4.6 $\times$ 10$^{-12}$(300/T)$^{0.64}$ & \citet{1995MNRAS.272...41S}\\
H${_2^+}$ + e $\rightarrow$ H + H  &   2.3 $\times$ 10$^{-8}$(300/T)$^{0.4}$ &\citet{1977JPhB...10.3797A}\\
H${_3^+}$ + e $\rightarrow$ H${_2}$ + H  &   2.9 $\times$ 10$^{-8}$(300/T)$^{0.65}$ & \citet{1994Sci...263..785S}\\
H${_3^+}$ + e $\rightarrow$ H + H + H  &   8.6 $\times$ 10$^{-8}$(300/T)$^{0.65}$ & \citet{1995PhRvL..74R4099D}\\
HeH${^+}$ + e $\rightarrow$ He + H &   1.0 $\times$ 10$^{-8}$(300/T)$^{0.6}$ & \citet{1989PhRvA..40.4318Y}\\
H + e $\rightarrow$ H${^+}$ + e + e &   2.91 $\times$ 10$^{-8}$ ($\rm\frac{1}{0.232+U}$)U$^{0.39}$exp(-U), U=13.6/$E_e$eV & \citet{1997ADNDT..65....1V}\\
He + e $\rightarrow$ He${^+}$ + e + e &   1.75 $\times$ 10$^{-8}$ ($\rm\frac{1}{0.180+U}$)U$^{0.35}$exp(-U),U=24.6/$E_e$eV& \citet{1997ADNDT..65....1V} \\
H + He${^+}$ $\rightarrow$ H${^+}$ + He &   1.25 $\times$ 10$^{-15}$(300/T)$^{-0.25}$ & \citet{2007ApJ...666....1G}\\
H${^+}$ + He $\rightarrow$ H + He${^+}$ &   1.75 $\times$ 10$^{-11}$(300/T)$^{0.75}$exp(-128000/T)& \citet{2007ApJ...666....1G}\\
O + e $\rightarrow$ O${^+}$ + e + e &   3.59 $\times$ 10$^{-8}$ ($\rm\frac{1}{0.073+U}$)U$^{0.34}$exp(-U),U=13.6/$E_e$eV & \citet{1997ADNDT..65....1V}\\
C + e $\rightarrow$ C${^+}$ + e + e &   6.85 $\times$ 10$^{-8}$ ($\rm\frac{1}{0.193+U}$)U$^{0.25}$exp(-U),U=11.3/$E_e$eV& \citet{1997ADNDT..65....1V}\\
O${^+}$ + e $\rightarrow$ O + $h\nu$ &   3.25 $\times$ 10$^{-12}$(300/T)$^{0.66}$ & \citet{2007AA...466.1197W}\\
C${^+}$ + e $\rightarrow$ C + $h\nu$ &   4.67 $\times$ 10$^{-12}$(300/T)$^{0.60}$ &  \citet{2007AA...466.1197W}\\
C${^+}$ + H $\rightarrow$ C + H${^+}$ &   6.30 $\times$ 10$^{-17}$(300/T)$^{-1.96}$exp(-170000/T)& \citet{1998ApJ...502.1006S}\\
C + H${^+}$ $\rightarrow$ C${^+}$ + H &     1.31 $\times$ 10$^{-15}$(300/T)$^{-0.213}$ & \citet{1998ApJ...502.1006S}\\
C + He${^+}$ $\rightarrow$ C${^+}$ + He &     2.50 $\times$ 10$^{-15}$(300/T)$^{-1.597}$ & \citet{2007ApJ...666....1G}\\
O${^+}$ + H $\rightarrow$ O + H${^+}$ &    5.66 $\times$ 10$^{-10}$(300/T)$^{-0.36}$exp(8.6/T)&  \citet{2007AA...466.1197W}\\
O + H${^+}$ $\rightarrow$ O${^+}$ + H &     7.31 $\times$ 10$^{-10}$(300/T)$^{-0.23}$exp(-226.0/T) & \citet{2007AA...466.1197W}\\
N + e $\rightarrow$ N${^+}$ + e + e &    4.82 $\times$ 10$^{-8}$ ($\rm\frac{1}{0.0652+U}$)U$^{0.42}$exp(-U),U=14.5/$E_e$eV & \citet{1997ADNDT..65....1V}\\
N${^+}$ + e $\rightarrow$ N + $h\nu$ &    3.46 $\times$ 10$^{-12}$(300/T)$^{0.608}$ & \citet{1973AA....25..137A}\\
Si + e $\rightarrow$ Si${^+}$ + e + e &   1.88 $\times$ 10$^{-7}$ ($\rm\frac{1+\sqrt{U}}{0.376+U}$)U$^{0.25}$exp(-U),U=8.2/$E_e$eV & \citet{1997ADNDT..65....1V}\\
Si${^+}$ + e $\rightarrow$ Si + $h\nu$ &    4.85 $\times$ 10$^{-12}$(300/T)$^{0.60}$ & \citet{1973AA....25..137A}\\
Si${^+}$ + e $\rightarrow$ Si${_2^+}$ + e + e &    6.43 $\times$ 10$^{-8}$ ($\rm\frac{1+\sqrt{U}}{0.632+U}$)U$^{0.25}$exp(-U),U=16.4/$E_e$eV &\citet{1997ADNDT..65....1V} \\
Si${_2^+}$ + e $\rightarrow$ Si${^+}$ + $h\nu$ &    1.57 $\times$ 10$^{-11}$(300/T)$^{0.786}$ & \citet{1973AA....25..137A}\\
H${^+}$ + Si $\rightarrow$ H + Si${^+}$ &   7.41 $\times$ 10$^{-11}$(300/T)$^{-0.848}$ & \citet{2007ApJ...666....1G}\\
He${^+}$ + Si $\rightarrow$ He + Si${^+}$ &     3.30 $\times$ 10$^{-9}$ & \citet{2007AA...466.1197W}\\
C${^+}$ + Si $\rightarrow$ C + Si${^+}$ &     2.10 $\times$ 10$^{-9}$ & \citet{2007AA...466.1197W}\\
H + Si${_2^+}$ $\rightarrow$ H${^+}$ + Si${^+}$ &    2.20 $\times$ 10$^{-9}$(300/T)$^{-0.24}$ & \citet{1996ApJS..106..205K}\\
H${^+}$ + Si${^+}$ $\rightarrow$ H + Si${_2^+}$ &    7.37 $\times$ 10$^{-10}$(300/T)$^{-0.24}$ & \citet{1996ApJS..106..205K}\\
\hline
\end{longtable}

\subsection{Radiative transfer} \label{subsec:raditran}
In order to calculate the absorption of stellar Ly$\alpha$ by the atmosphere of the planets, the radiative transfer equations are as following
\begin{equation}
F_{out}=F_{in}e^{-\tau}
\end{equation}

\begin{equation}
\tau=\int_{Z_{0}}^{Z} n\sigma dl
\end{equation}

Finally, the absorption depth by the planet and its atmosphere can be expressed as
\begin{equation}
Absorption\,depth=\frac{F_{in}-F_{out}}{F_{in}}
\end{equation}


In equation (3)-(5), F$_{in}$ is the incident intrinsic stellar flux (note that it is different from stellar Fxuv.) F$_{out}$ is the emergent flux due to the occultation by planets and the absorption by their surrounding atmospheres along the ray path. In the radiation transfer equations, the factor $\tau$ represents the optical depth, which is dependent on the atmospheric structure and directly pertaining to the particles' number density n and cross section $\sigma$.

\subsection{Cross section---Ly$\alpha$ }\label{subsubsec:crosslh}
Ly$\alpha$ absorption occurs in a hydrogen atom when an electron absorbs 10.2 ev energy and jumps from n=1 level to n=2 level, where n is the quantum number. The cross-section of Ly$\alpha$ absorption can be evaluated via,
\begin{equation}
\sigma_{12}=\frac{\pi e^2f_{12}\phi_{\nu}}{m_ec}
\end{equation}
in $\rm cm^2$, where e is the elementary charge of an electron, m$_e$ is the electronic mass, $f_{12}$ is the oscillator strength, which is 0.4162 at 1215.67 $\rm\AA$ \citep{Mihalas1978}. $\phi_\nu$ is the Voigt profile, which combines Doppler and Lorentz profiles. The Voigt profile is related to Voigt function H(a,u) through the following equations \citep{Rybicki 2004},

\begin{equation}
\phi_\nu=(\Delta\nu_D)^{-1}\pi^{-\onehalf}H(a,u)
\end{equation}
\begin{equation}
H(a,u)\equiv\frac{a}{\pi}\int\limits_{-\infty}^{+\infty}\frac{e^{-y^2}dy}{a^2+(u-y)^2}
\end{equation}
\begin{equation}
a\equiv\frac{\Gamma}{4\pi\Delta\nu_D}, u\equiv\frac{\nu-\nu_0}{\Delta\nu_D}
\end{equation}

\begin{equation}
\Gamma=\gamma+2\nu_{col}, \Delta\nu_D=\frac{\nu_0}{c}\sqrt{\frac{2kT}{m_a}}
\end{equation}

where a is the damping parameter and u is the frequency offset, $\nu_0$ is the line center frequency, $\Delta\nu_D$ (assuming no turbulence) is the Doppler width, and $\Gamma$ is the transition rate. Here we set the damping parameter a to be $4.699\times10^{-4}$.

\section{\textbf{Results}} \label{sec:result2}

\subsection{\textbf{The mass loss rates}\label{result21}}
\subsubsection{\textbf{The dependence of $\dot{M}$ on Fxuv and $\overline{\rho}$}\label{result21}}
We investigated 442 systems in which the number of Jupiter-like, Saturn-like, Neptune-like and Earth-like planets are 151, 78, 104 and 109. The mass loss rate predicted by the hydrodynamic model is defined as:
\begin{equation}
\dot{M}=4\pi r^{2}\rho \upsilon.
\end{equation}
here $\rho$ and $\upsilon$ are the density and the velocity of the escaping particles.

In addition, the absorbed XUV irradiation is converted into heat and does work on the particles of the atmosphere to overcome the gravitational potential and supply the particles' kinetic and thermal energy. According to \citet{2007A&A...472..329} and \citet{2009AA...506..399}, the mass loss rates can be expressed as
\begin{equation}
\rm\dot{M}=\frac{\pi F_{xuv} \eta R_{xuv}^2}{\Delta\phi+\frac{\upsilon^2_{R_L}-\upsilon^2_{R_0}}{2}+ c_p(T_{R_L}-T_{R_0})}
\end{equation}
where R$_{xuv}$ is the XUV absorption radius (at which the mean optical depth is 1). $\eta$ is the heating efficiency, which is the ratio of the gas heating energy to the whole XUV energy input. $\Delta\phi$, $\upsilon^2_{R_L}-\upsilon^2_{R_0}$ and $c_p(T_{R_L}-T_{R_0}$) are the variations (from the lower atmosphere boundary to the Roche Lobe) of the gravitational potential, kinetic and thermal energy respectively. $R_0$ and $R_L$ are the locations of the atmosphere lower boundary and the Roche Lobe. $\upsilon$ and T are the velocity and temperature of the particles. $c_p$ is the specific heat at a constant pressure per unit mass. In the energy-limited loss approximation \citep{2003APJL...598..L121, 2007A&A...472..329,2009AA...506..399}, the kinetic and thermal energy of the escaping particles are far smaller than the gravitational potential. Thus, the energy-limited equation is expressed as
\begin{equation}
\dot{M}= \frac{\pi  F_{xuv} \eta R_{xuv}^2}{\Delta\phi}
\end{equation}
When the effect of stellar tidal force is included, $\Delta\phi = \frac{G M_P}{R_P} K(\xi) $ \citep{2007A&A...472..329}. Therefore the energy-limited equation can be further expressed as
\begin{equation}
\rm\dot{M}=\frac{3\beta_{xuv}^{2}\eta Fxuv}{4 K(\xi) G\rho},
\end{equation}
where $\beta_{xuv}$ is the ratio of XUV absorption radiu to planetary radius.(Here we use the subscript xuv to distinguish it from the spectral index $\beta$.) $\rho$ is the planetary mean density and G is the gravitational constant. K($\xi$) is the potential energy reduction factor due to the stellar tidal forces  \citep{2007A&A...472..329},
\begin{equation}
K(\xi)=1-\frac{3}{2\xi}+\frac{1}{2\xi^3}
\end{equation}
with
\begin{equation}
\xi=(\frac{Mp}{3M_\star})^{\frac{1}{3}}\frac{a}{Rp}
\end{equation}
where M$_{p}$ and M$_\star$ are the planetary and stellar masses, a is the separation between planet and its host star, R$_{p}$ is the planetary radius. By comparing Equation (12) and Equation (14), one can found that the energy-limited formula can be revised by the terms of the kinetic and thermal energy. In the paper, the Equation (12)is defined as revised energy-limited formula.  We will discuss the effect of the kinetic and thermal energy in Section 3.1.2.

The energy-limited equation hints that the mass loss rates are the function of the XUV flux and the mean density.
Figure. 2 shows the dependence of the mass loss rate on the properties of exoplanet and the XUV flux. First, it is clear that the mass loss rates increase with the increase of the XUV irradiation as expected. The mass loss rates of Jupiter-like planets are in the order of magnitude of 10$^{9}$-10$^{10}$ g/s when the XUV integrated flux is about 200 erg/cm$^{2}$/s. In the case of F$_{xuv}$=4$\times$10$^{5}$ erg/cm$^{2}$/s, the mass loss rates of Jupiter-like planets increase a factor of $\sim$ 1000, which is in the order of magnitude of 10$^{13}$ g/s. Such a trend is also found for the Saturn-like, Neptune-like and Earth-like planets. For example, the mass loss rates of the Earth-like planets vary almost linearly with F$_{xuv}$ from 10$^{8}$ g/s to 10$^{11}$ g/s. Second, we note that the mass loss rates of the planets decrease with the decrease of their sizes if the integrated flux is given. For the Saturn-like planets, the mass loss rates decline by a factor of a few comparing with those of Jupiter-like planets. However, for those smaller planets the mass loss rates can decrease by an order of magnitude or more. Such a behaviour reflects the fact that the atmospheric escape of the planet is inverse with the mean density.

\begin{figure}
\centering
\includegraphics[width=4in,height=3.in]{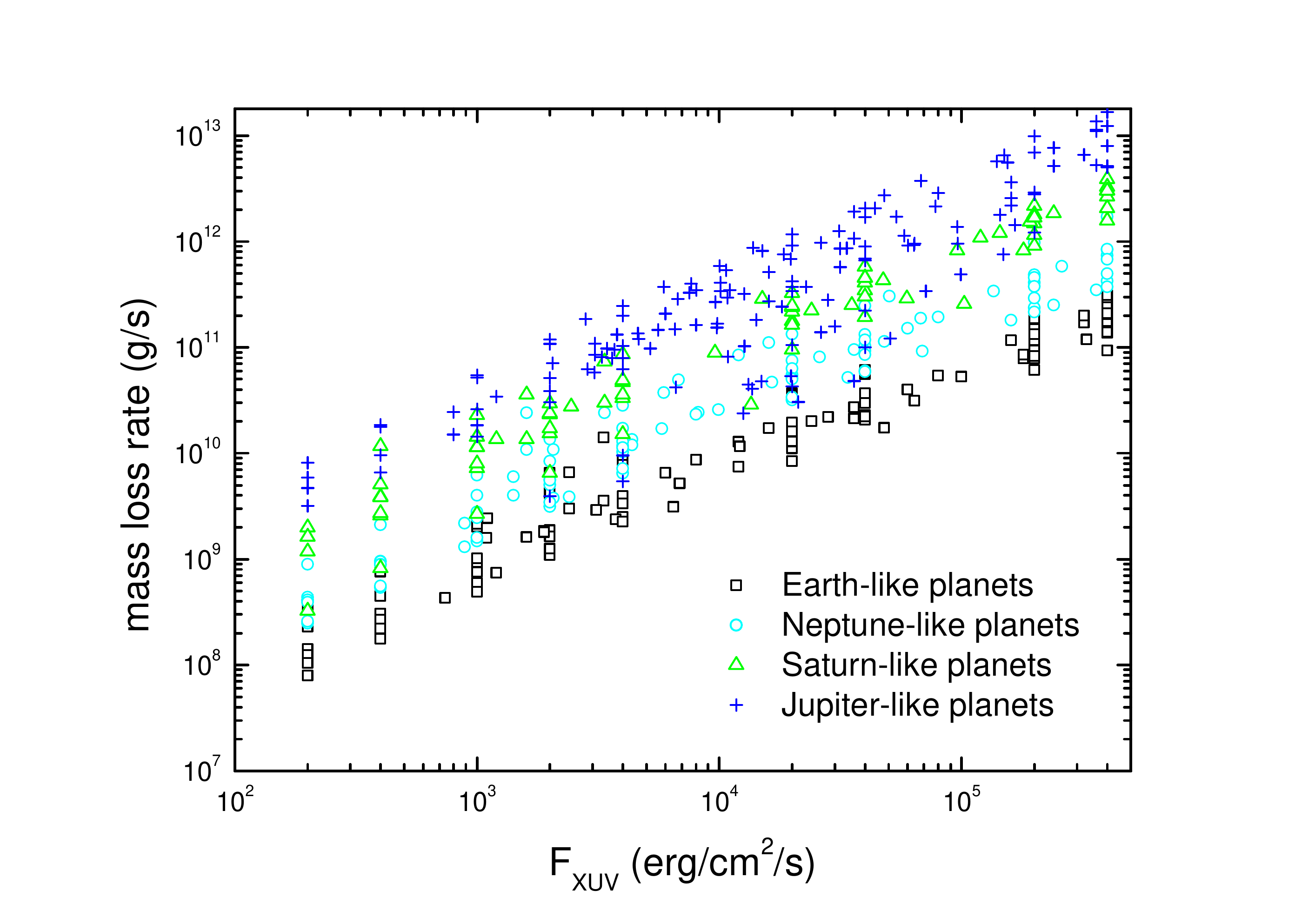}
\caption{The mass loss rates of planets at different XUV fluxes. Cross:  Jupiter-like planets. Triangles:  Saturn-like planets. Circles:  Neptune-like planets. Squares:  Earth-like planets. \label{fdtdmass} }
\end{figure}

To clarify, we selected some planets of our sample to calculate the mass loss rates on the different XUV fluxes. The corresponding value of the fluxes are 200, 400, 1000, 2000, 4000, 2$\times$10$^{4}$, 4$\times$10$^{4}$, 2$\times$10$^{5}$ and 4$\times$10$^{5}$ erg cm$^{-2}$ s$^{-1}$ . We showed the correlation between the mean density and the mass loss rate in Figure. 3. Note that the real flux in the calculation is divided by a factor of 4 so that the mass loss rates of a few Jupiter-like planets can not be calculated in the situation of the low XUV flux. The mass loss rates of those planets are calculated by using the energy-limited equation (The value of $\beta_{XUV}^2\eta$ is obtained from our fitting formulae. For details, see Section. 4.1.). Evidently, the mass loss rates of the planets with lower density are higher than those of the planet with higher density. Thus, the hydrodynamic simulation confirmed the physical validity of the energy-limited assumption.
\begin{figure}
\centering
\includegraphics[width=4in,height=3.in]{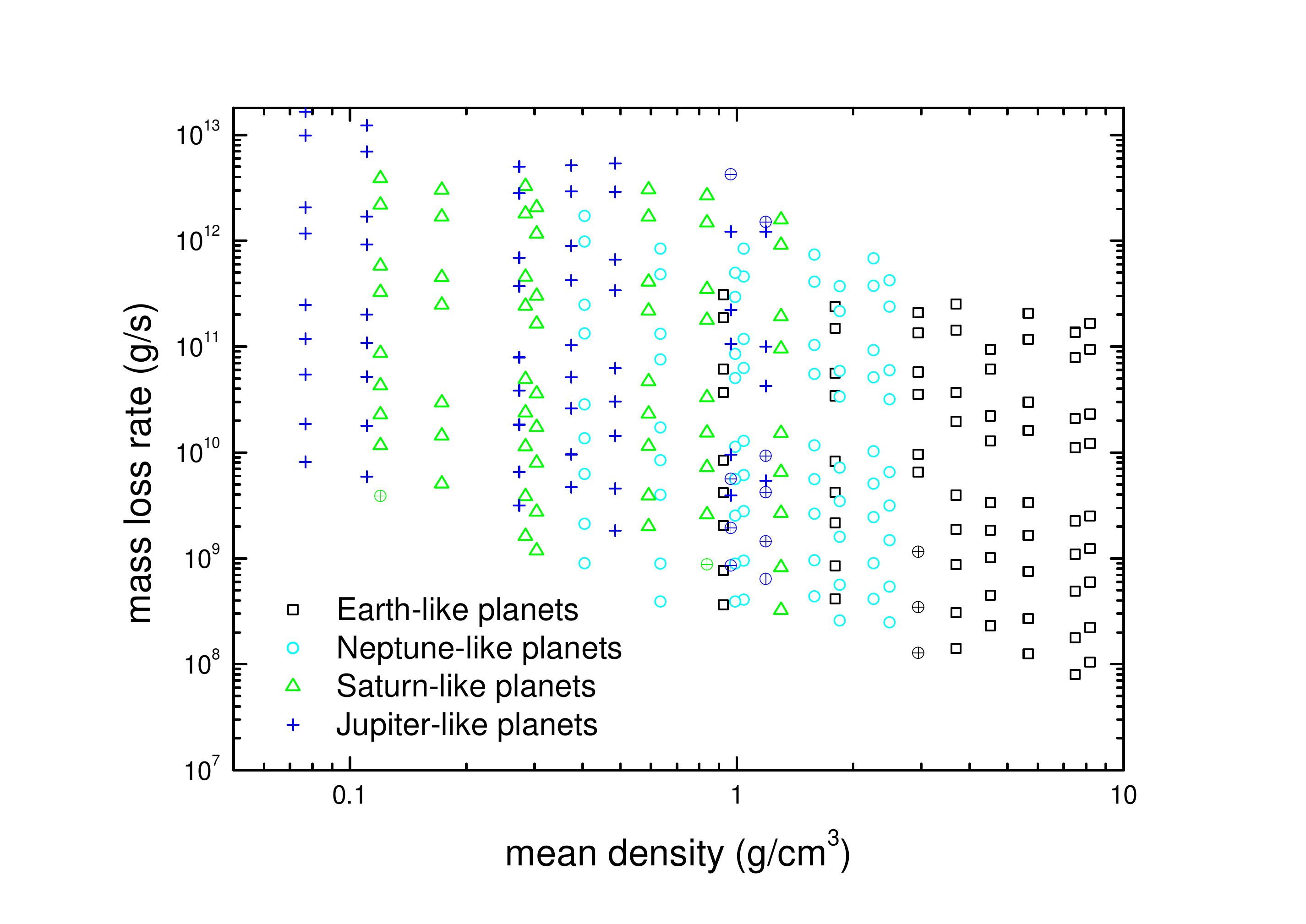}
\caption{The mass loss rates of planets with different density. Cross:  Jupiter-like planets. Triangles:  Saturn-like planets. Circles:  Neptune-like planets. Squares:  Earth-like planets. $\oplus$: the mass loss rates calculated by using the energy-limited equation. \label{fdtdmass} }
\end{figure}

However, we note that the mass loss rates of different planets are slightly different even if the densities of the planets are same. For example, at $\overline{\rho}$ $\approx$ 0.28 g cm$^{-3}$ the mass loss rates of Jupiter-like planets are higher than those of Saturn-like planets although the XUV fluxes are same. Such behaviours also occur at $\overline{\rho}$ $\approx$ 1 g cm$^{-3}$ and $\overline{\rho}$ $\approx$ 1.8 g cm$^{-3}$.
This reflects that the energy available for driving the escape of the atmosphere are different for different planets. There are three factors that can result in the variations of the mass loss rate. First, the heating efficiency of different planets is different. Second, the effective area of energy deposition are different for different planets. Third, the tidal forces of the host stars. The effect of tidal force has been discussed by \citet{2007A&A...472..329} who found that the tidal forces of star can enhance the mass loss rates. Thus, we will discuss the first two factors below.

\subsubsection{\textbf{The heating efficiency and the XUV absorption radius }\label{result21}}
A main factor in determining the mass loss rates is the heating fraction of the XUV radiation. The net heating efficiency $\eta$ is defined as:

\begin{equation}
\eta=(H_{heat}-L_{cooling})/\Sigma_{\nu} F \nu
\end{equation}

where $F\nu$ is the input XUV energy at frequency $\nu$,  H$_{heat}$ and L$_{cooling}$ are the radiative heating and cooling, respectively.
We calculated the heating efficacy of our $\sim$ 400 planets and found that the heating efficiencies are insensitive to an individual parameter, but it is dependent of the product of the XUV flux and the gravitational potential (hereafter log(F$_{xuv}GM_p/R_p$)). The left panel of Figure. 4. shows the heating efficacy with respect to log(F$_{xuv}GM_p/R_p$). The triangles represent the planets with gravitational potential lower than 1.5$\times$10$^{13}$ erg g$^{-1}$ and the cross (plus sign) are planets with gravitational potential higher than 1.5$\times$10$^{13}$ erg g$^{-1}$. For planets with gravitational potential smaller than 1.5$\times$10$^{13}$ erg g$^{-1}$, we can see from this figure that the larger the log(F$_{xuv}GM_p/R_p$), the higher the heating efficiency $\eta$. For the planets that concentrate on the range of 14 $<$ log(F$_{xuv}GM_p/R_p$) $<$ 16, the heating efficiencies of most planets are lower than 0.3. Only a few planets appear higher heating efficiency. When 16 $<$ log(F$_{xuv}GM_p/R_p$) $<$ 18, the heating efficiency rises from about 0.13 to 0.45. For planets with log(F$_{xuv}GM_p/R_p$) larger than 18, $\eta$ varies in the range of 0.3-0.45. The appearance of $\eta$ is, however, different for the planets with relatively large gravitational potential (the cross in Figure. 4 ). In fact, the gravitational potential makes a separation to the  $\eta$. We can see from the left panel of Figure .4 that the heating efficiency is generally lower for the planets with very large gravitational potential compared to those with relatively small gravitational potential. We further showed the dependence of $\eta$ on the gravitational potential in the right panel of Figure. 4. For the planets with relatively small gravitational potential, the heating efficiency can vary a factor of 9 and appear an increasing trend with the increase of gravitational potential. At the same time, the values of $\eta$ express a transition around $\sim$ 1$\times$10$^{13}$ erg g$^{-1}$ above which the heating efficiency decrease with the gravitational potential. Such behavior causes the lower mass loss rates for some Jupiter-like planets. As shown in Figure. 2, we found that there are some Jupiter-like planets of which the mass loss rates deviate from the general trend. They are generally smaller than the mass loss rates of other Jupiter-like planets and even smaller than those of some Neptune-like and Earth-like planets. The mass loss rates are related to the mean densities. For such planets, their mean densities are around 1 g/cm$^3$. The higher mean densities result in the decrease of a factor of a few for the mass loss rates comparing with those of Jupiter-like planet with lower mean densities. At the same time, these planets all have relatively large gravitational potentials ($>$ 1.5$\times$10$^{13}$ erg/g) which leads to the relatively small heating efficiency $\eta$ (most are in the range of 0.05-0.1). Therefore, for these planets the mass loss rates could be about 10 times lower than that of other Jupiter-like planets. We also note that \citet{2016A&A...585..2s} found a rapid decrease of the evaporation efficiencies (By assuming heating efficiency $\eta$=1 in calculating the mass loss rates predicted by the energy-limited equation, they defined evaporation efficiency $\eta_{eva}=\frac{\dot{M}_{model}}{\dot{M}_{enengy-limited}}$ and found $\eta$=1.2 $\eta_{eva}$ .) when the gravitational potential is higher than $\sim$ 1.3$\times$10$^{13}$ erg g$^{-1}$, which is similar with our results. However, we did not find the linear dependence of $\eta$ on the gravitational potential when the values of gravitational potential are lower than $\sim$ 1.3$\times$10$^{13}$ erg g$^{-1}$ (Figure. 2 of \citet{2016A&A...585..2s}). Our calculations based on a larger sample suggest that the heating efficiency can vary a factor of a few even if the gravitational potentials of the planets are same.


\begin{figure}
\begin{minipage}[t]{0.5\linewidth}
\centering
\includegraphics[width=3.6in,height=3.0in]{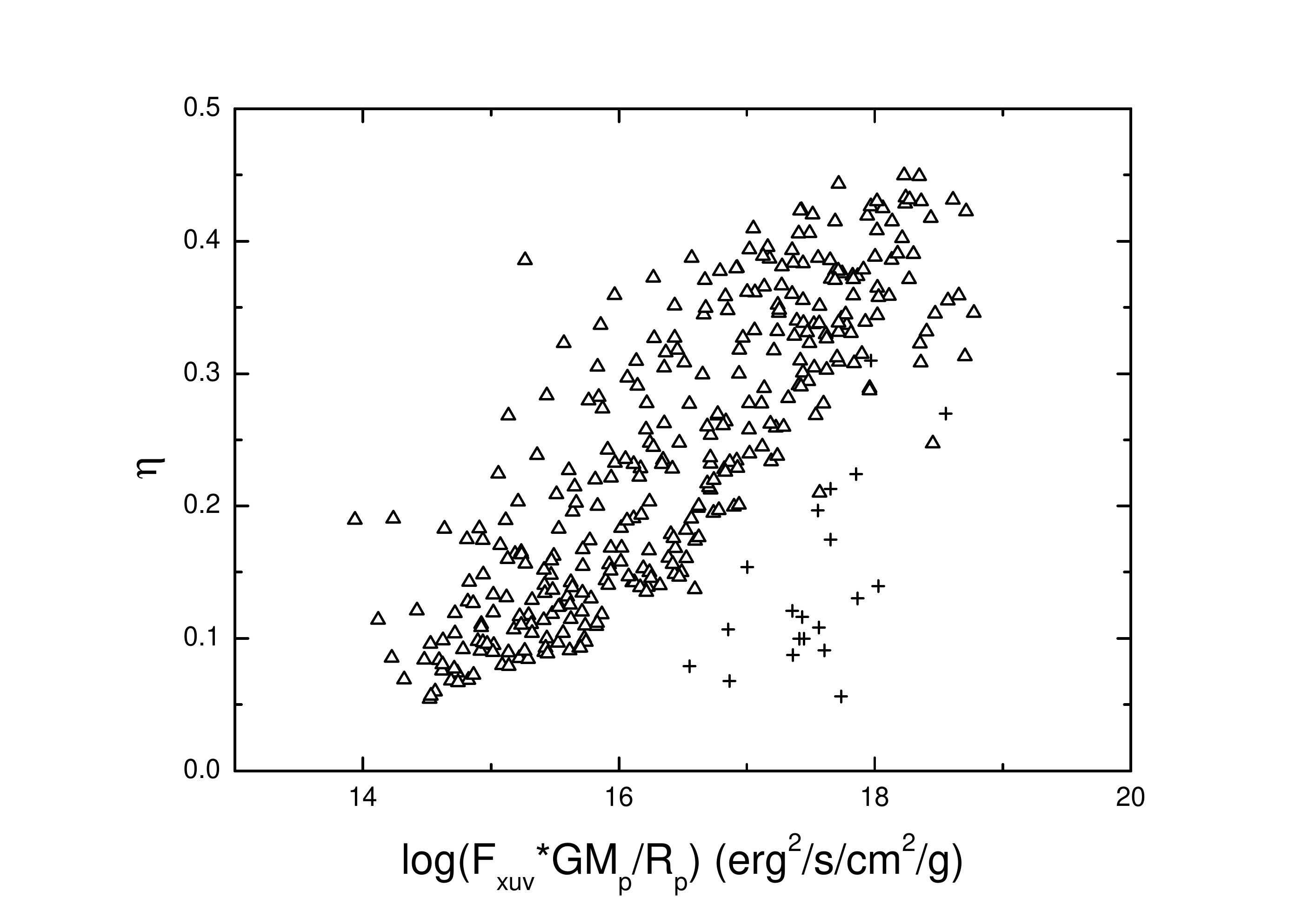}
\end{minipage}
\begin{minipage}[t]{0.5\linewidth}
\centering
\includegraphics[width=3.6in,height=3.0in]{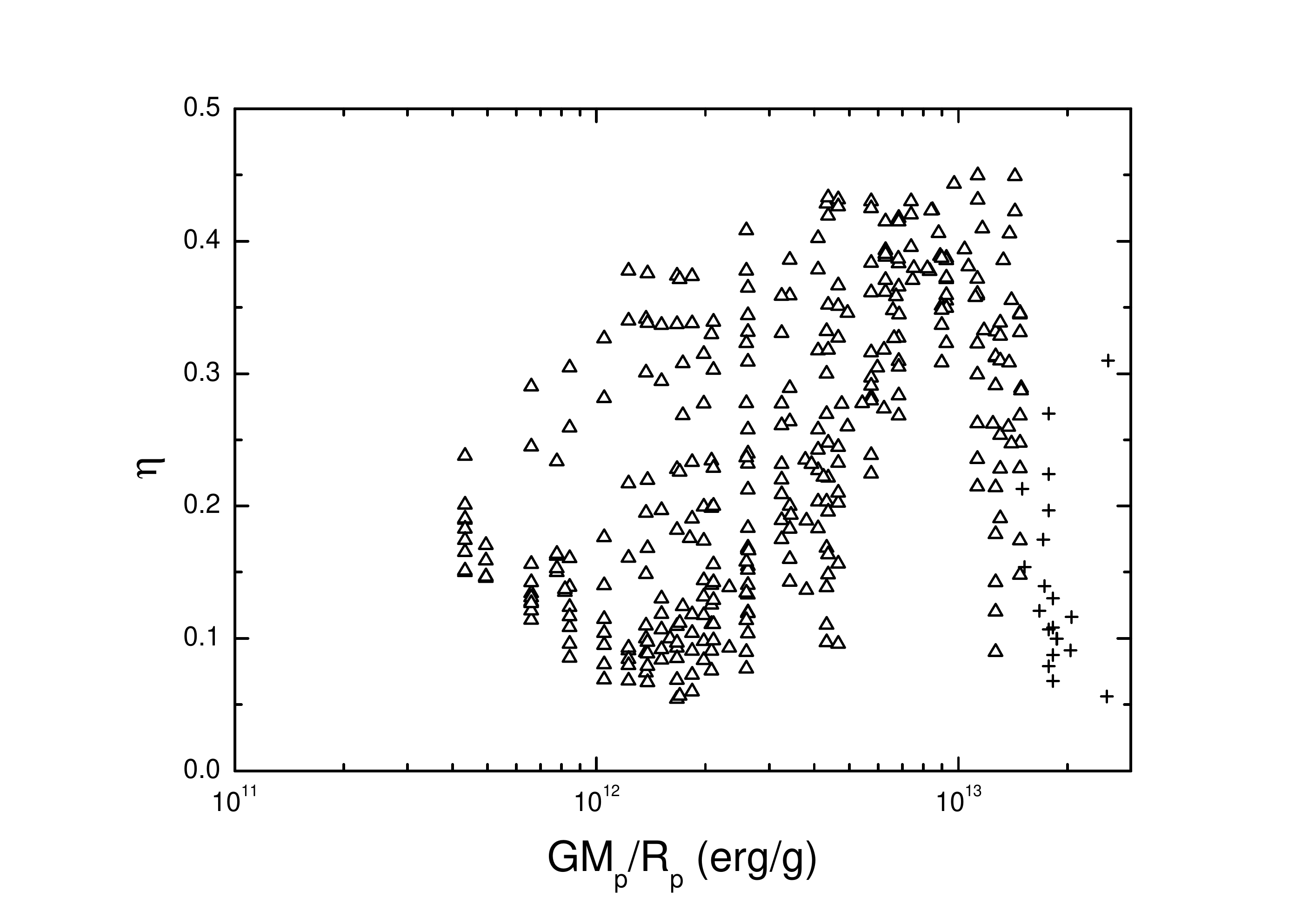}
\end{minipage}
\caption{Left panel: The relation between heating efficiency and the log(F$_{xuv}GM_p/R_p$). Right panel: The relation between heating efficiency and the $GM_p/R_p$. Triangles: the planets with a gravitational potentials are lower than 1.5$\times$10$^{13}$ erg g$^{-1}$.  Cross:  the planets with a gravitational potentials are greater than 1.5$\times$10$^{13}$ erg g$^{-1}$.}
\end{figure}

Furthermore, one needs to know the mean XUV absorption radius R$_{XUV}$ in order to calculate the mass loss rates predicted by the energy-limited equation. Here R$_{XUV}$ is defined as:
\begin{equation}
R_{XUV}=\frac{\sum_{\nu} R_{\nu}(\tau_{\nu}=1) F_{\nu}}{F_{XUV}}.
\end{equation}
In Equation (16), R$_{\nu}(\tau_{\nu}=1)$ is the radius where the optical depth at frequency $\nu$ is unit. F$_{\nu}$ is the XUV flux at frequency $\nu$. We calculated the XUV absorption radius by equation (16) for all the planets. The variations of R$_{XUV}$ are shown in Figure. 5.
The upper panel is the  R$_{XUV}$ with respect to the planetary radii. For all the planets,  R$_{XUV}$ is in the range of 1.05-1.7 R$_{p}$. We found that the values of R$_{XUV}$ are related to the the sizes of planets. For the Jupiter-like planets, the R$_{XUV}$ mainly concentrates on a small range which is 1.1-1.2 R$_{p}$. The values of R$_{XUV}$ vary from 1.1 to 1.5  R$_{p}$ when the sizes of planets are in the range of 0.2-0.8 R$_{J}$. For planets with radii less than 0.2 R$_{J}$, the R$_{XUV}$ could reach 1.4-1.7 R$_{p}$. Compared to the Earth-like planets, the R$_{XUV}$ of larger planets such as Jupiter-like planets could decrease by a factor of 1.5.

The middle panel is the R$_{XUV}$ with respect to the planetary mass. The R$_{XUV}$ is basically negatively correlated to the planetary mass. For the planets with masses less than 0.1 M$_{J}$, the R$_{XUV}$ tends to be larger, especially for those less than 0.01 M$_{J}$. When the masses become higher than 0.2 M$_{J}$, the R$_{XUV}$ is mainly in 1.1-1.2 R$_p$. Both the upper panel and the middle panel suggest that the smaller planets such as Earth-like planets tend to have larger R$_{XUV}$, which means their XUV absorption radii are relatively far from the planetary surface. The lower panel of Figure. 5 expresses how the R$_{XUV}$ varies with the gravitational potential. It is clear that the absorption radii increase with the decrease of the gravitational potentials. For the planets with a high gravitational potential, the values of R$_{XUV}$ are in the range of 1.1-1.2 R$_p$. The increase of R$_{XUV}$ is dramatic when the gravitational potentials are lower than $\sim$ 10$^{12}$ erg g$^{-1}$. This is reasonable because for these planets the gravitational potentials are so low that the atmospheres can expand to higher altitudes to absorb the stellar XUV irradiation.

\begin{figure}
\centering
\includegraphics [width=4.in,height=2.5in] {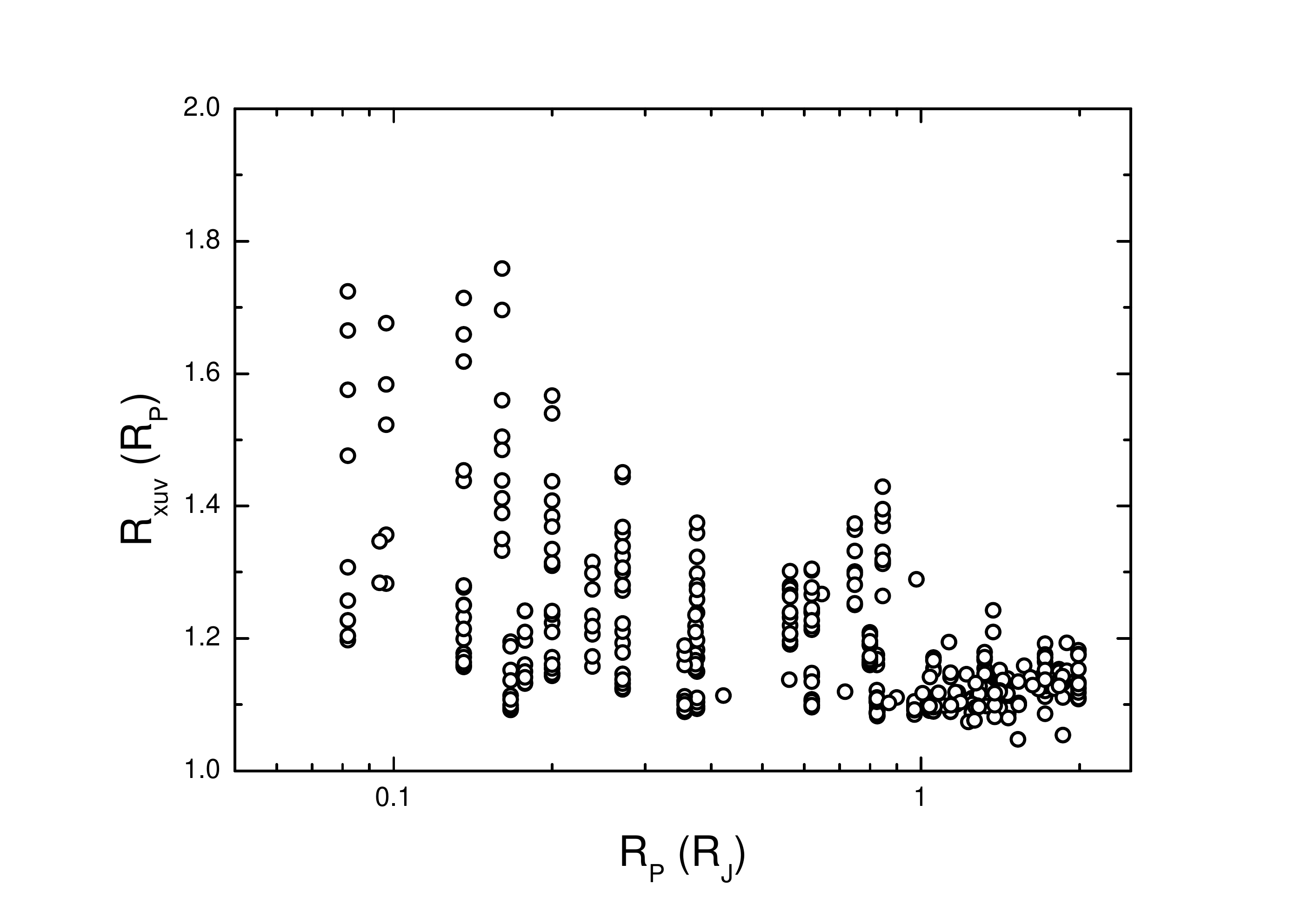}
\centering
\includegraphics [width=4.in,height=2.5in] {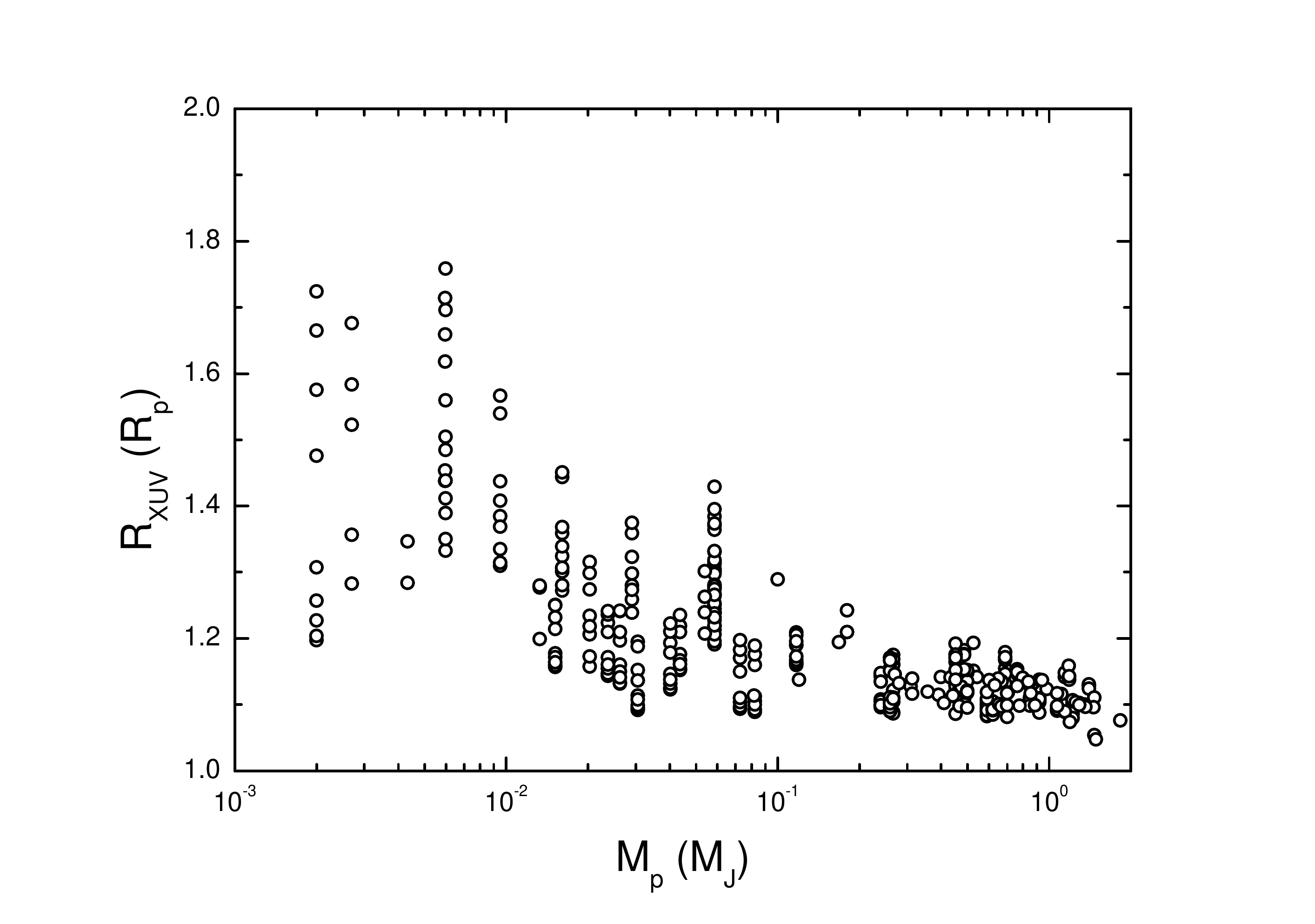}
\centering
\includegraphics [width=4.in,height=2.5in] {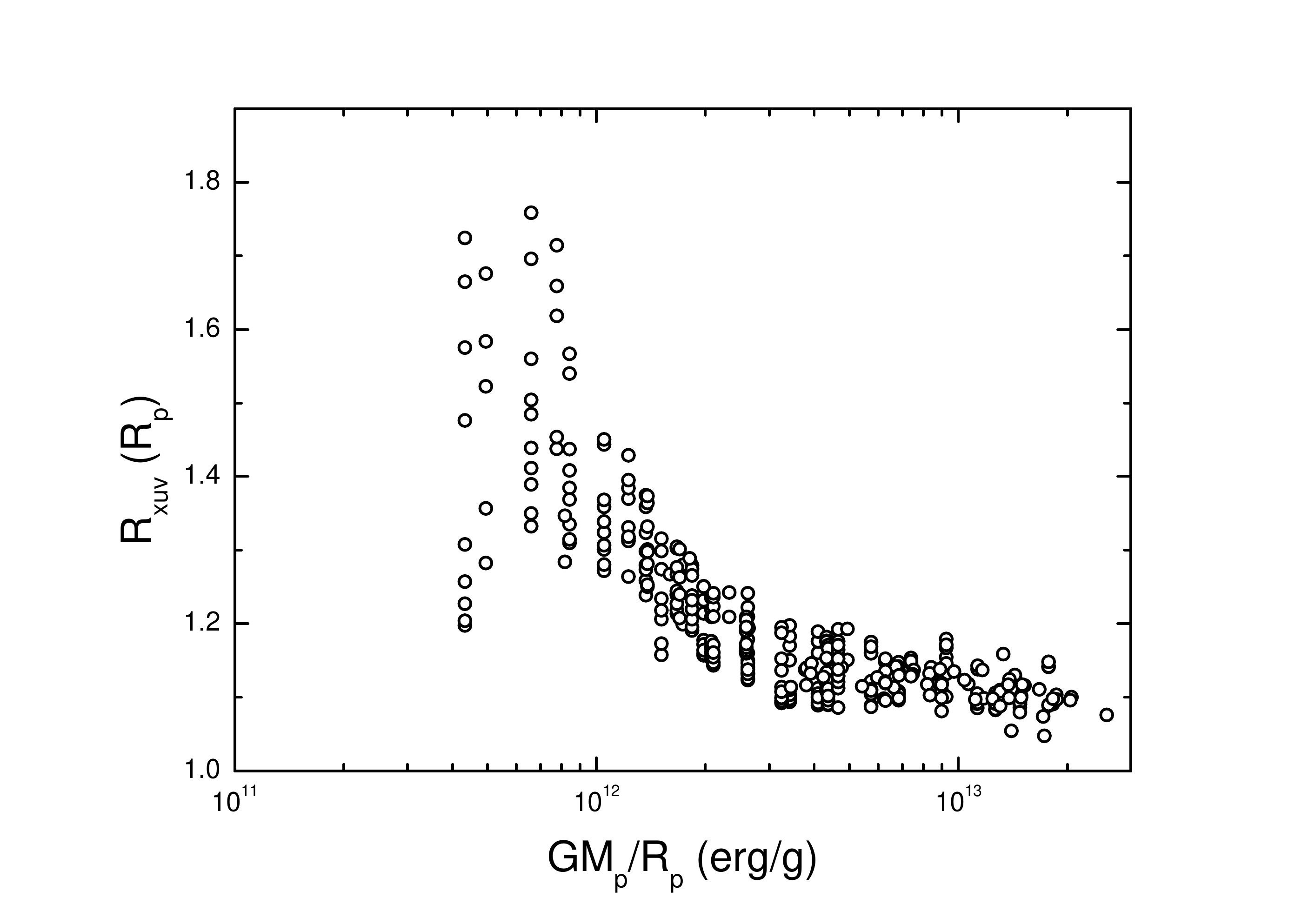}
\caption{The upper panel: the variations of R$_{XUV}$ with the radii. The middle panel: the variations of R$_{XUV}$ with the masses. The lower panel shows the dependence of R$_{XUV}$ on the gravitational potential.}
\end{figure}

In Figure. 6a and 6b we compared the mass loss rates predicted by our hydrodynamic model(y-axis) and the Equation (12) and (14)(x-axis). The mass loss rates of Equation (12) and (14) are calculated by using our $\beta_{XUV}$ and $\eta$. The mass loss rates in Figure. 6b are corrected by the kinetic and thermal energy of the escaping atmosphere (Equation. 12). Different symbols represent the mass loss rates of different types of planets. Before correcting for the kinetic and thermal energy of the escaping atmosphere (see Figure .6a), the hydrodynamic mass loss rates are basically consistent with that of the energy-limited when the mass loss rates are lower than a ``critical value" for different type of planets. For Earth-like and Neptune-like planets, it is about 10$^{10}$-10$^{11}$g/s. For Saturn-like and Jupiter-like planets, it is about 10$^{11}$-10$^{12}$g/s. By comparing with Figure. 2, we found that the corresponding XUV levels of the critical mass loss rates are about 2$\times$10$^4$-3$\times$10$^4$ erg/cm$^2$/s for Earth-like and Neptune-like planets and $\sim$ 4$\times$10$^4$ erg/cm$^2$/s for Saturn-like and Jupiter-like planets. Above the XUV radiation level, the hydrodynamic mass loss rates of the planets are lower than the energy-limited ones, especially for the smaller planets such as Earth-like and Neptune-like planets. The deviation of the two kinds of mass loss rates comes from the kinetic and thermal energy of the escaping atmosphere. To specify, we show the influence of the kinetic and thermal energy in Figure. 6b. After correcting for the kinetic and thermal energy of the escaping atmosphere, the mass loss rates of our model are generally consistent with that predicted by the revised energy-limited equation (Equation (12)) for all kinds of planets.

The uncertainties of the mass loss rates obtained from the energy-limited formula are mainly attributed to the unknown heating efficiencies and the absorption radii of the XUV irradiation. Many studies have applied the fixed heating efficiencies for some types of planets. In our study, the heating efficiencies are different for different planets even if they are same types of planets. At the same time, the absorption radii also vary with the physical parameters of planets (Figure. 5). In Figure. 6a and 6b, we used the heating efficiencies and the absorption radii of our hydrodynamic models to remove the uncertainties. However, the deviations are still evident for the energy-limited case. This highlights the importance of kinetic and thermal energy in using the energy-limited equation. As shown in Figure. 6a and 6b, a portion of the energy of XUV irradiation is converted the kinetic and thermal energy of the escaping atmosphere. Neglecting the kinetic and thermal energy will result in an overestimation to the mass loss rates when the energy-limited formula is used. We further show the dependence of the kinetic and thermal energy to the planetary parameters in the Figure. 6c. It is clear that the sum of the kinetic and thermal energy increases with the increase of log(F$_{xuv}GM_p/R_p$). The increasing trend is obvious for all types of planets although there is a relatively large spread for Earth-like planets and Jupiter-like planets. Finally, We show the ratios of the kinetic and thermal energy to the gravitational potential in Figure. 6d. The ratios are smaller than unit when log(F$_{xuv}GM_p/R_p$) is smaller than 16. With the increase of log(F$_{xuv}GM_p/R_p$), the ratios become greater than unit for most of Earth-like planets and Neptune-like planets while the ratios are still smaller than unit for most of Jupiter-like planets. The ratios of Saturn-like planets express a mixed variation for the case of high log(F$_{xuv}GM_p/R_p$). Same trend also appears in the Figure. 6a. The mass loss rates of Jupiter-like planets predicted by the energy-limited formula are more consistent with that of hydrodynamic model because the gravitational potential is dominant. For smaller planets, the deviation of the mass loss rate becomes obvious with the increase of the mass loss rate. Thus, the results obtained from the energy-limited formula should be revised by the kinetic and thermal energy of the escaping atmosphere, especially for those planets with high F$_{xuv}GM_p/R_p$.

\begin{figure}
\begin{minipage}[t]{0.5\linewidth}
\centering
\includegraphics[width=3.6in,height=3.0in]{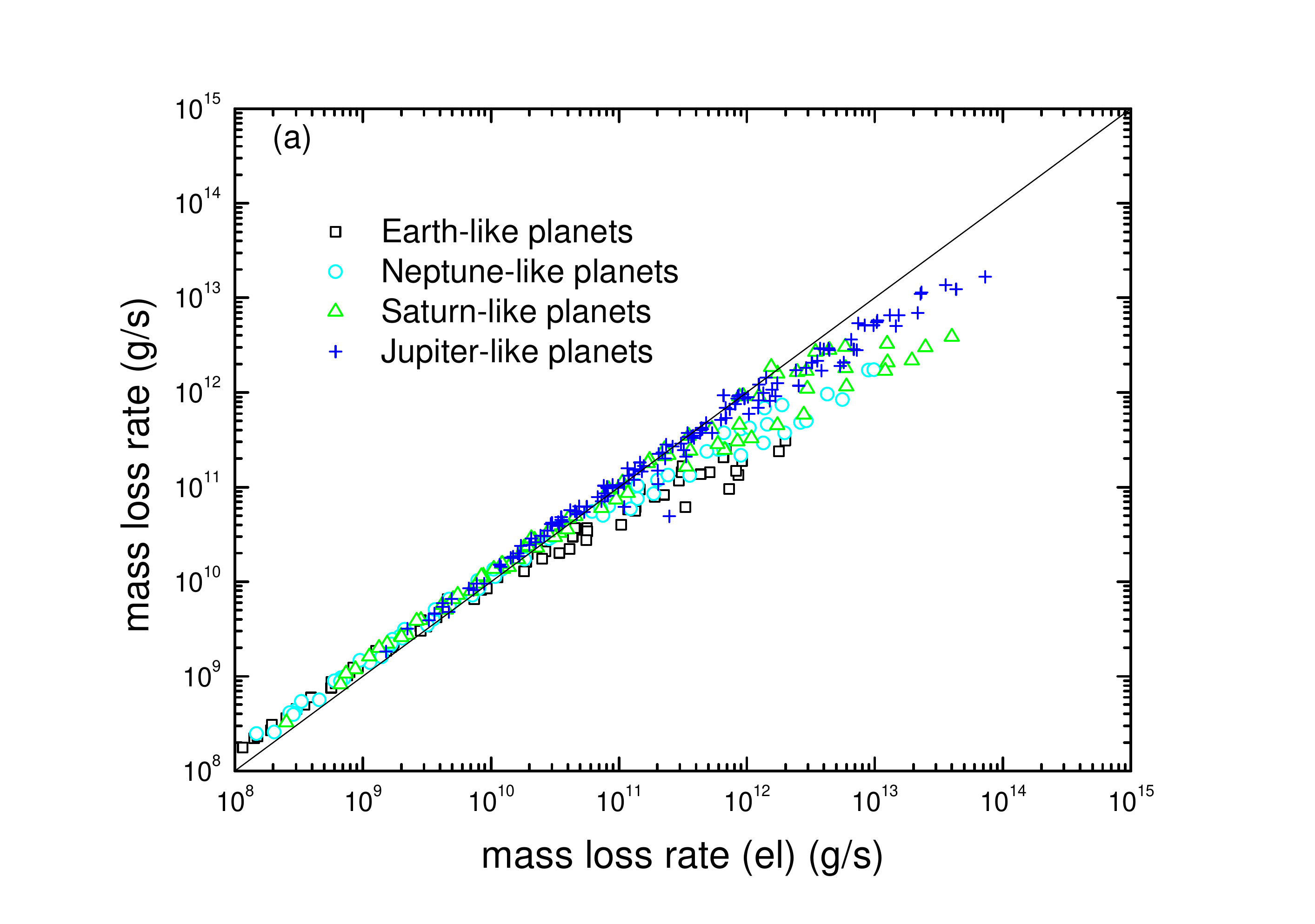}
\end{minipage}
\begin{minipage}[t]{0.5\linewidth}
\centering
\includegraphics[width=3.6in,height=3.0in]{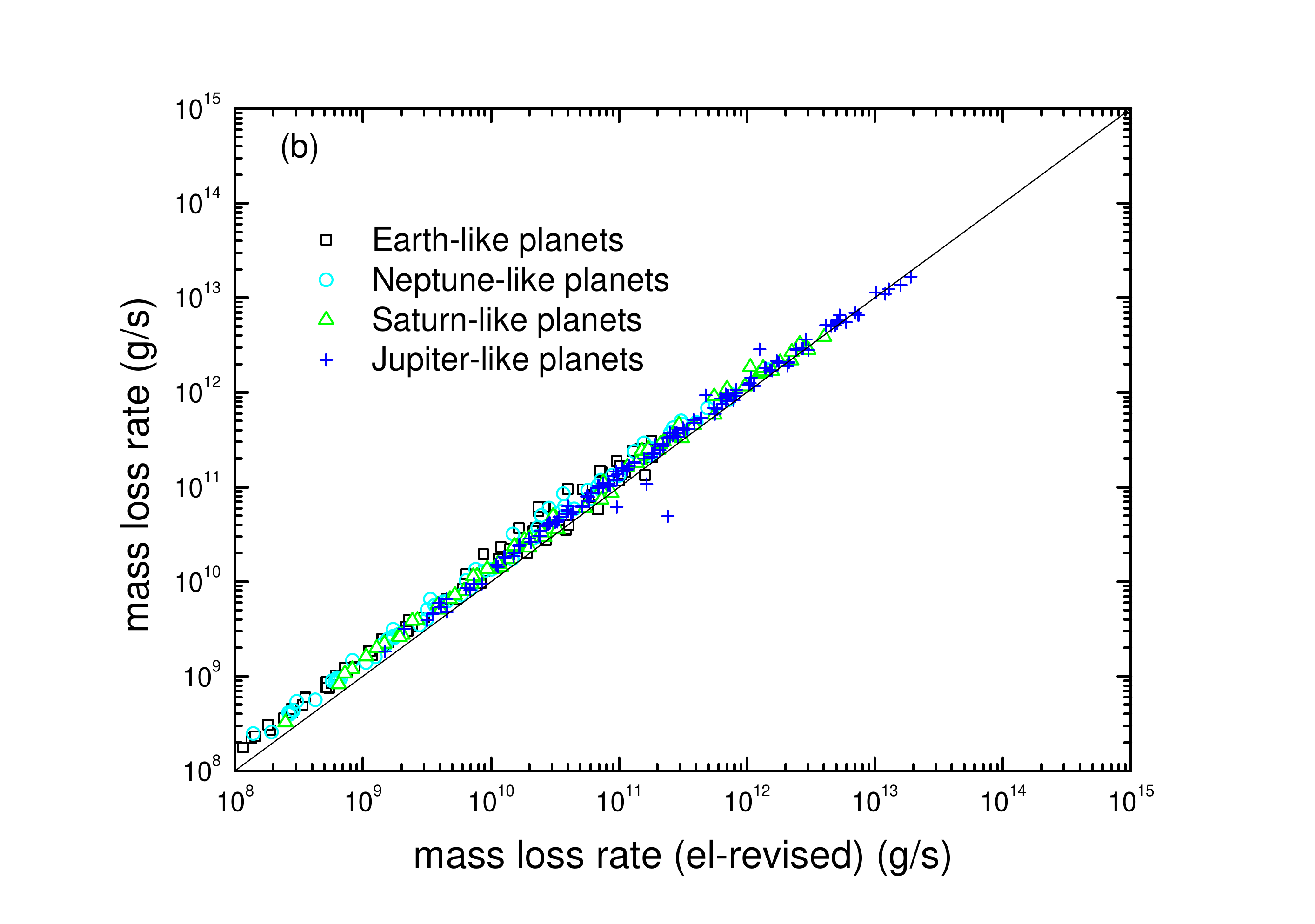}
\end{minipage}
\begin{minipage}[t]{0.5\linewidth}
\centering
\includegraphics[width=3.6in,height=3.0in]{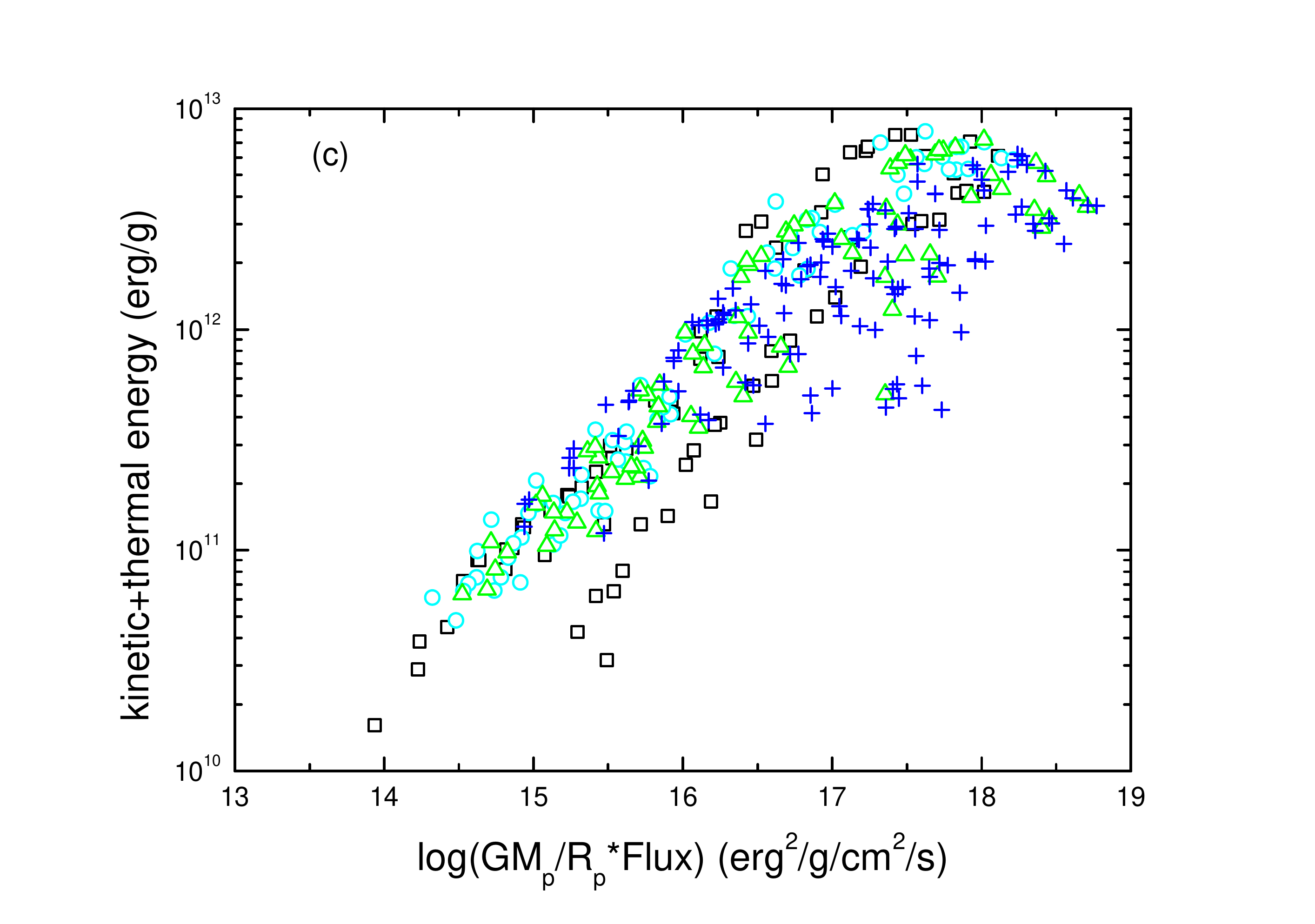}
\end{minipage}
\begin{minipage}[t]{0.5\linewidth}
\centering
\includegraphics[width=3.6in,height=3.0in]{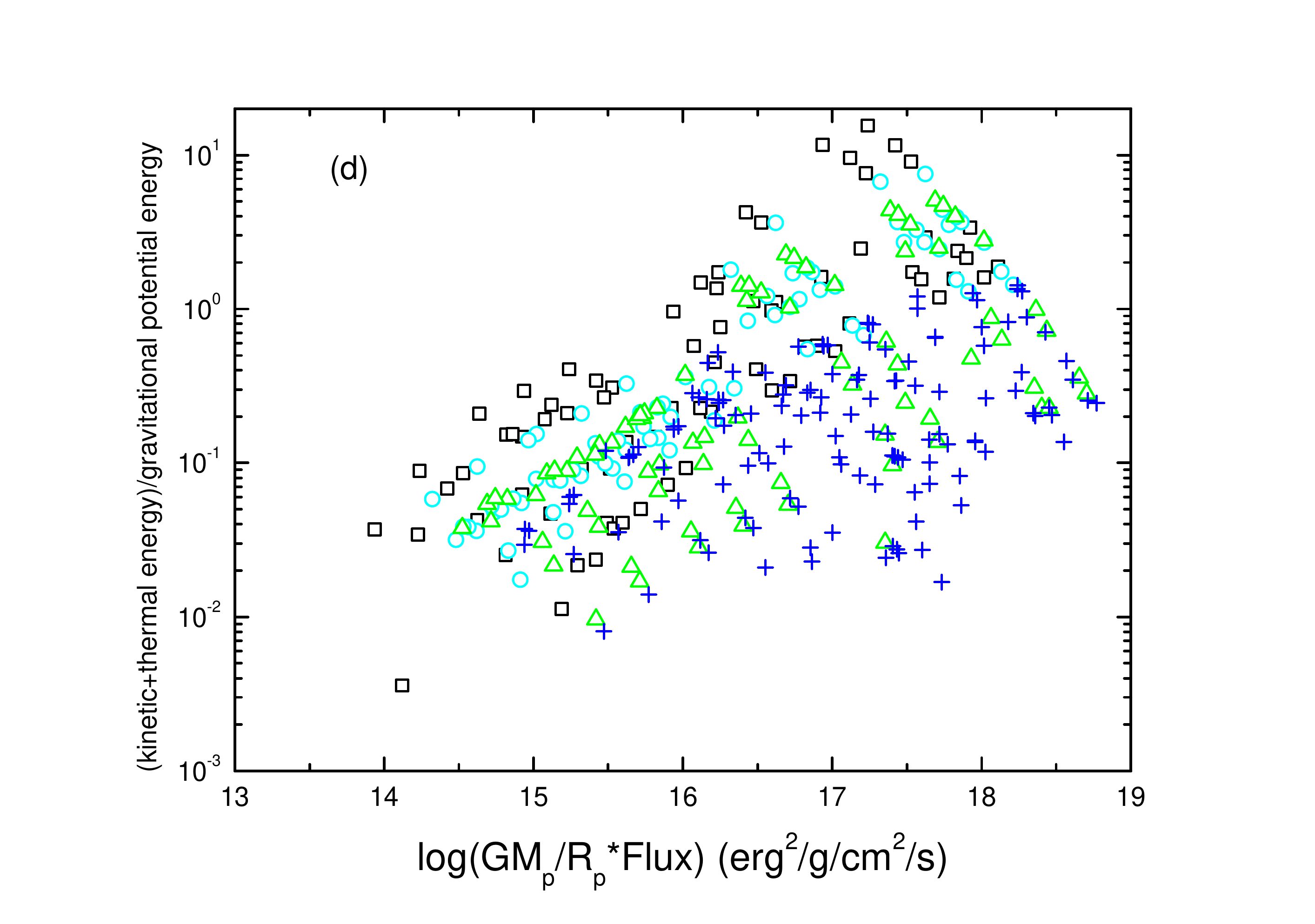}
\end{minipage}
\caption{Panel (a) and (b) show the comparison of the mass loss rates predicted by our model and that of energy-limited equation.  The mass loss rates of energy-limited assumption are calculated by using our $\beta_{XUV}$ and $\eta$. In the panel (a) $\dot{M}$(el) is the mass loss rates calculated by the energy-limited equation and is not corrected for the kinetic and thermal energy of the escaping atmosphere. In the panel (b) $\dot{M}$(el-revised) is corrected for the kinetic and thermal energy of the escaping atmosphere. The panel (c) expresses the sum of the kinetic and thermal energy. The ratios of he kinetic and thermal energy to the gravitational potential is shown in panel (d).}
\end{figure}

\subsection{\textbf{The dependence of the absorption depth of Lyman $\alpha$ on F$_{XUV}$ and $\overline{\rho}$}}\label{result21}

We calculated the absorption of stellar Ly$\alpha$ by using our sample and found that the excess absorption levels of Ly$\alpha$ could be higher if the planets have lower mean densities and relatively higher integrated Fxuv. To specify, Figure \ref{fdtdmass} shows the absorption depths in the Fxuv-$\rho$ diagram. The x-axis is the planetary mean density and the y-axis is the integrated Fxuv. The absorption depths are divided into three or four levels depending on the absorption depth range. The absorptions of different depth are distinguished by different colored symbols. For the left and right panels of Figure. 7,  the velocity ranges are [-150, -50]$\cup$[50, 150] km/s and [-150, 150] km/s from the Ly$\alpha$ line center, respectively. In the [-50, 50] km/s range from the Ly$\alpha$ line center, the stellar Ly$\alpha$ could be contaminated by the ISM \citep{2019A&A...622..46} and the geocoronal Ly$\alpha$ emission lines, so we exclude the range in the left panel of Figure. 7.

\begin{figure}
\includegraphics[width=3.5in,height=2.85 in]{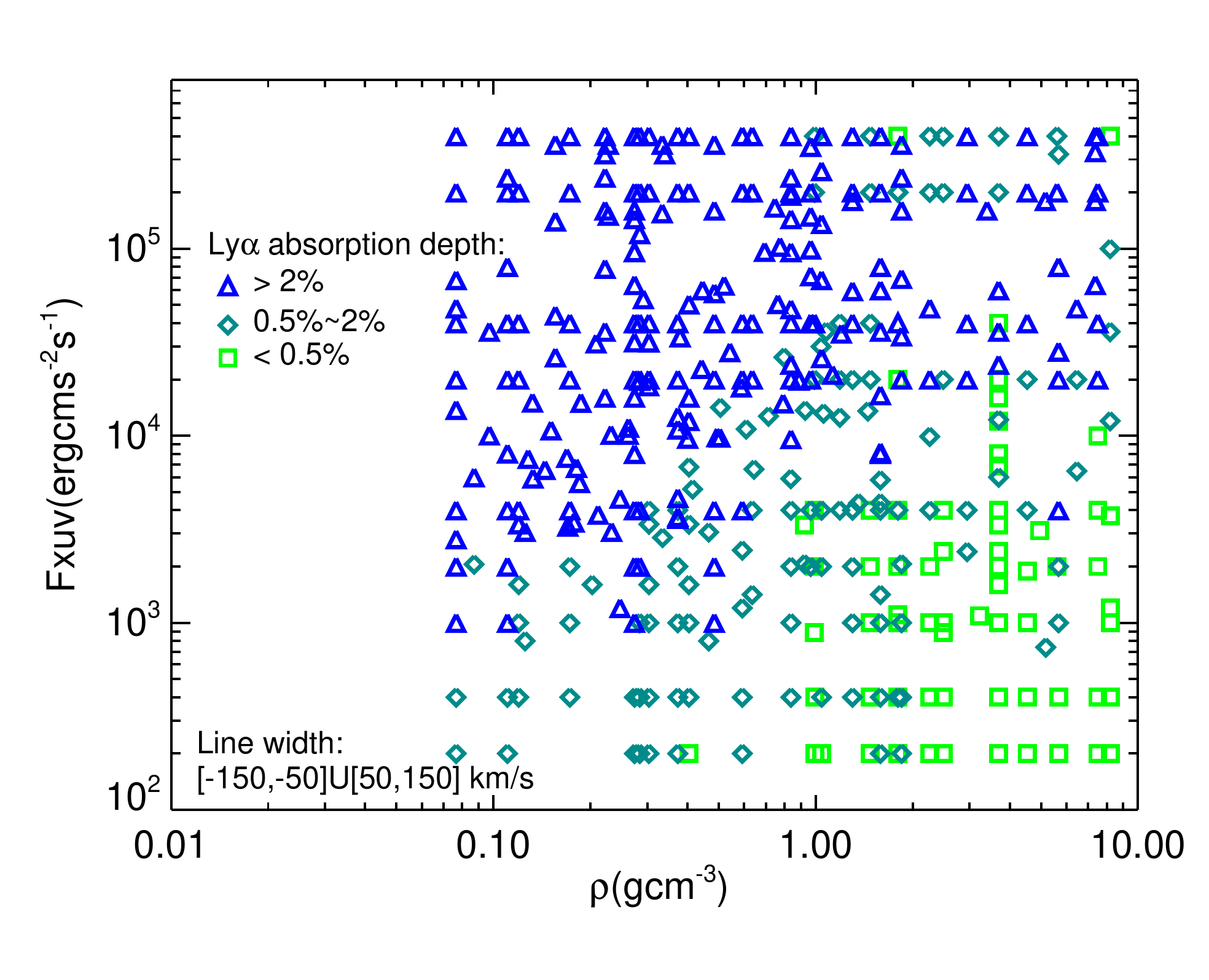}
\includegraphics[width=3.5in,height=3 in]{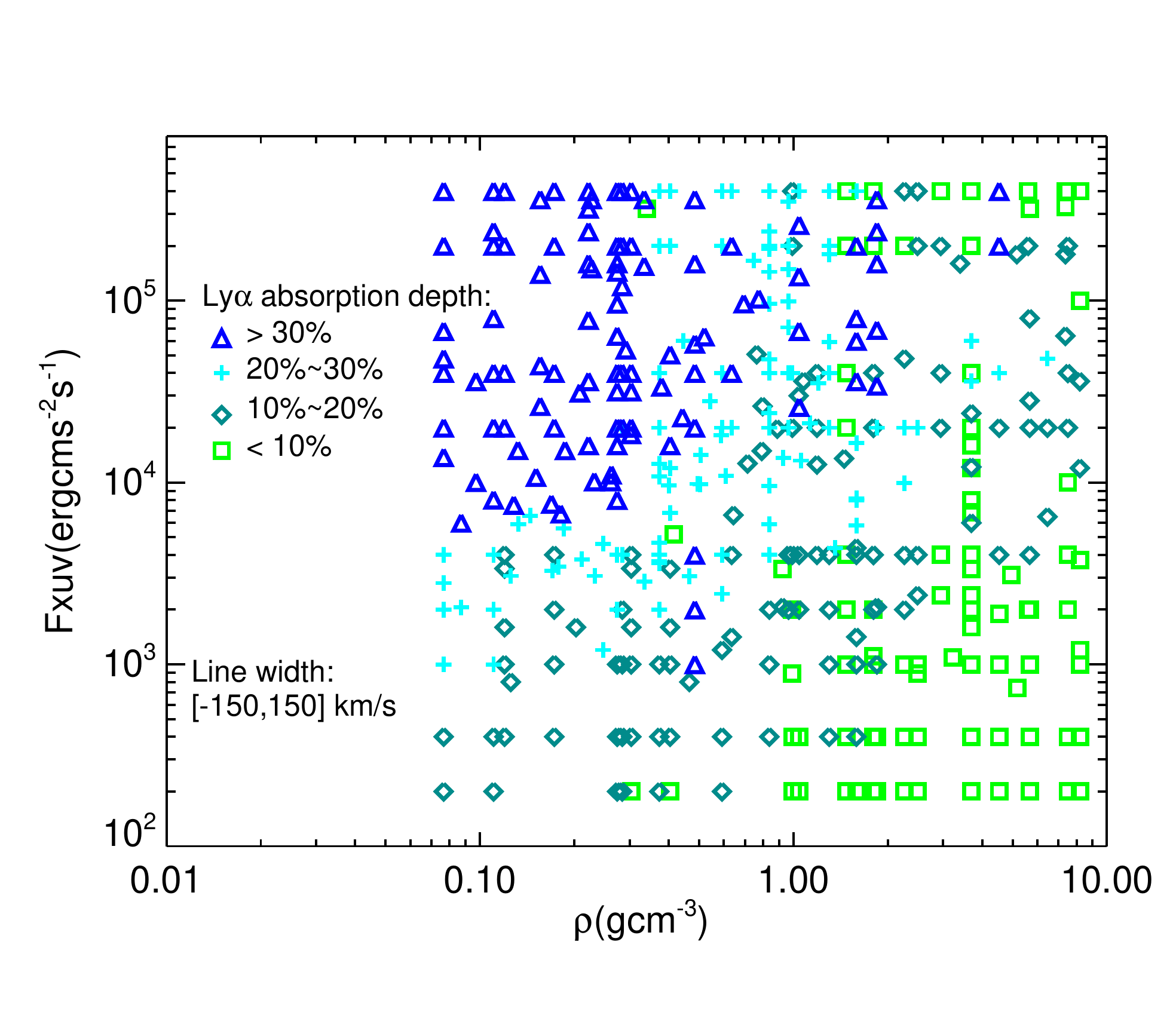}
\caption{The statistical distribution of absorption depths. The x-axis is the planetary mean density and y-axis is the integrated XUV flux. Different colored symbols represent different absorption levels. The left and right panels are the distributions of Ly$\alpha$ absorption depths (not including the optical occultation) which are calculated in [-150,-50]$\cup$[50,150] km/s and [-150,150]km/s from the line center 1215.67$\rm\AA$ respectively. \label{fdtdmass} }

\end{figure}

For the Ly$\alpha$ absorption depth, we can see from Figure. 7 that no matter what wavelength range it is, the average absorption levels have a similar distribution trend. The regions of stronger Ly$\alpha$ absorptions are in the upper left region of the Fxuv-$\rho$ diagram where the planets receive higher irradiation and have lower mean densities. This indicates higher mass loss rates around those planets. It is shown in the left panel of Figure. 7 that for planets with lower mean densities, the decreasing Fxuv can cause the decrease of the absorption levels. For those planets with medium or high mean density ( $>$ 1g/cm$^{3}$), the dependence of Ly$\alpha$ absorption on the XUV flux shows a middle sensibility. Comparing with those planets with lower densities, one can see that there is a mixed region in which the absorption levels are variable in a large range. For instance, in the case of F$_{XUV}$=10$^{4}$ erg/cm$^{2}$/s the absorptions produced by the atmosphere vary with the mean densities of the planets from $>$ 5\% and $<$ 1\%. Finally, the absorption levels are very low for those planets with high densities ($\rho > $ 1g/cm$^{3}$ ) and low XUV fluxes ( lower than 10$^{4}$ erg/cm$^{2}$/s). In the lower right region, almost all absorptions are lower than 1\%. Generally, those planets with higher densities are the Earth-like or Neptune-like planets.

It is also clear from the left panel of Figure. 7 that the absorptions are dependent of XUV flux. Most absorption levels of the planets are still higher than 1\% if the flux is higher than 2$\times$ 10$^{4}$erg cm$^{2}$/s. In the range of 10$^{3}$erg cm$^{2}$/s-10$^{4}$erg cm$^{2}$/s, the absorptions decrease with the increase of density. For those planets with high density, their absorptions can be a factor of a few smaller than those planets with lower density. Below 10$^{3}$erg cm$^{2}$/s, only planets with low densities appear significant absorptions. Such behaviors can also be found in the right panel of Figure. 7. Although the absorption of line center is included, the distributions of Ly$\alpha$ absorptions reflect the same trend as shown in the left panel of Figure. 7. The difference of two cases are that the absorption levels of the right panel are far lager than those of the left panel because of the strong absorption in the line center. \citet{2019A&A...623A.131K} modeled the in-transit absorption in Ly$\alpha$ for the terrestrial planets with nitrogen and hydrogen-dominated atmospheres under different levels of stellar irradiation. They also found the deeper absorption for the planets with higher XUV and smaller density, which is consistent with our results.
\begin{figure}
\centering
\includegraphics[width=3.6in,height=2.6in]{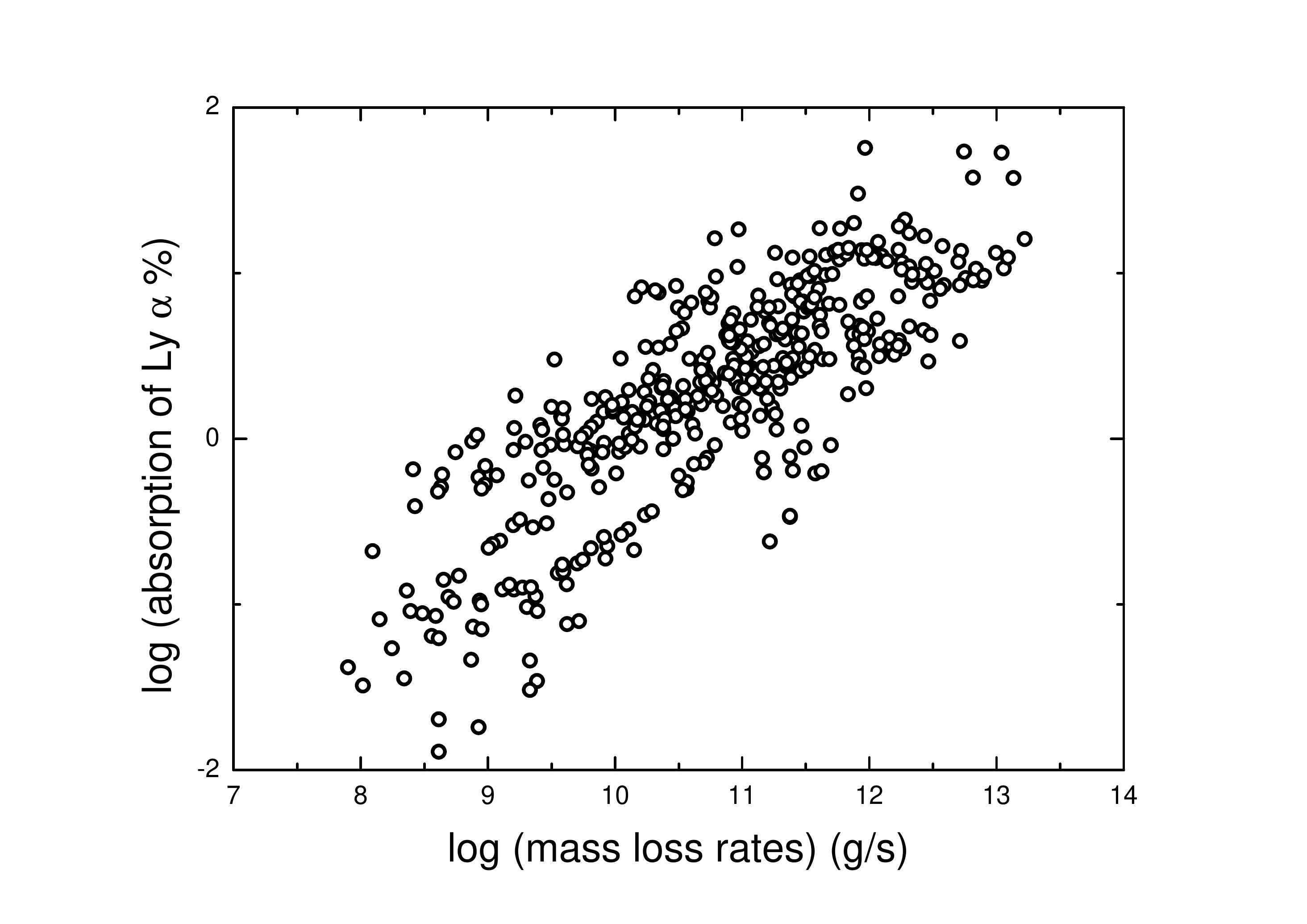}
\caption{The Ly$\alpha$ absorption in the range of [-150,-50]$\cup$[50,150] km/s as a function of mass loss rates.}
\end{figure}

The absorption levels are proportional to the mass loss rates. We showed the absorption of [-150,-50]$\cup$[50,150] km/s in Figure. 8. It is clear form Figure. 8 that higher absorption levels correspond to higher mass loss rates. For instance, the absorption depths of most planet can attain 1\% if the mass loss rates are higher than 10$^{10}g/s$. For the planets with the mass loss rates higher than 10$^{11}g/s$, the absorption level can attain 10\% or higher. In the cases of the mass loss rates lower than 10$^{10}g/s$, most planets appear low absorption level except a few planets. Finally, the absorption can be neglected if the mass loss rates are lower than 10$^{9}g/s$.

\section{Discussion} \label{sec:discu}

\subsection{\textbf{The fit to the parameters of the (revised) Energy-limited equation }}\label{subsec:result22}
The (revised) energy-limited equation is a convenient way to evaluate the planetary mass loss rates \citep{2003APJL...598..L121}. In fact, Figure. 2 and Figure. 3 can be explained at a certain extent by the energy-limited assumption. However, it is not easy to evaluate the planetary mass loss rates by using Equation (12) and Equation (14) because the values of $\beta_{xuv}$ and  $\eta$ must be specified. For convenience, one only needs to know the product of $\beta_{xuv}^{2}$ and  $\eta$ in evaluating the mass loss rates. Therefore, we fitted the parameters $\beta_{xuv}^{2}$ $\eta$ for the planet with the gravitational potential smaller than 1.5$\times$10$^{13}$ erg g$^{-1}$. The planets are classified into four categories by their gravitational potentials. We used different functions to fit the values of $\beta_{xuv}^{2}$ $\eta$ and showed the fit in Figure. 9. The fit of $\beta_{xuv}^{2}$ $\eta$ to log(GM$_{p}$F$_{xuv}$/R$_{p}$) can be expressed as :

\[\beta_{xuv}^{2} \eta=\begin{cases}
 7.980(\pm 2.315)-1.091(\pm 0.292) \theta +0.0383(\pm 0.0092) \theta^{2} & GM_{p}F_{xuv}/R_{p} < 1.5\times 10^{12}\\
 -1.901(\pm 0.0537)+0.135(\pm 0.0033) \theta & 1.5\times 10^{12} \leqslant GM_{p}F_{xuv}/R_{p} < 5\times 10^{12}\\
 -0.581(\pm 0.1218)+0.061(\pm 0.0072) \theta & 5\times 10^{12} \leqslant GM_{p}F_{xuv}/R_{p} < 1\times 10^{13}\\
 -1.232(\pm 0.0192)+0.093(\pm 0.0110) \theta & 1\times 10^{13} \leqslant GM_{p}F_{xuv}/R_{p} < 1.5\times 10^{13},
 \end{cases}\]
where $\theta$=log(GM$_{p}$F$_{xuv}$/R$_{p}$).

\begin{figure}
\centering
\includegraphics[width=3.6in,height=2.6in]{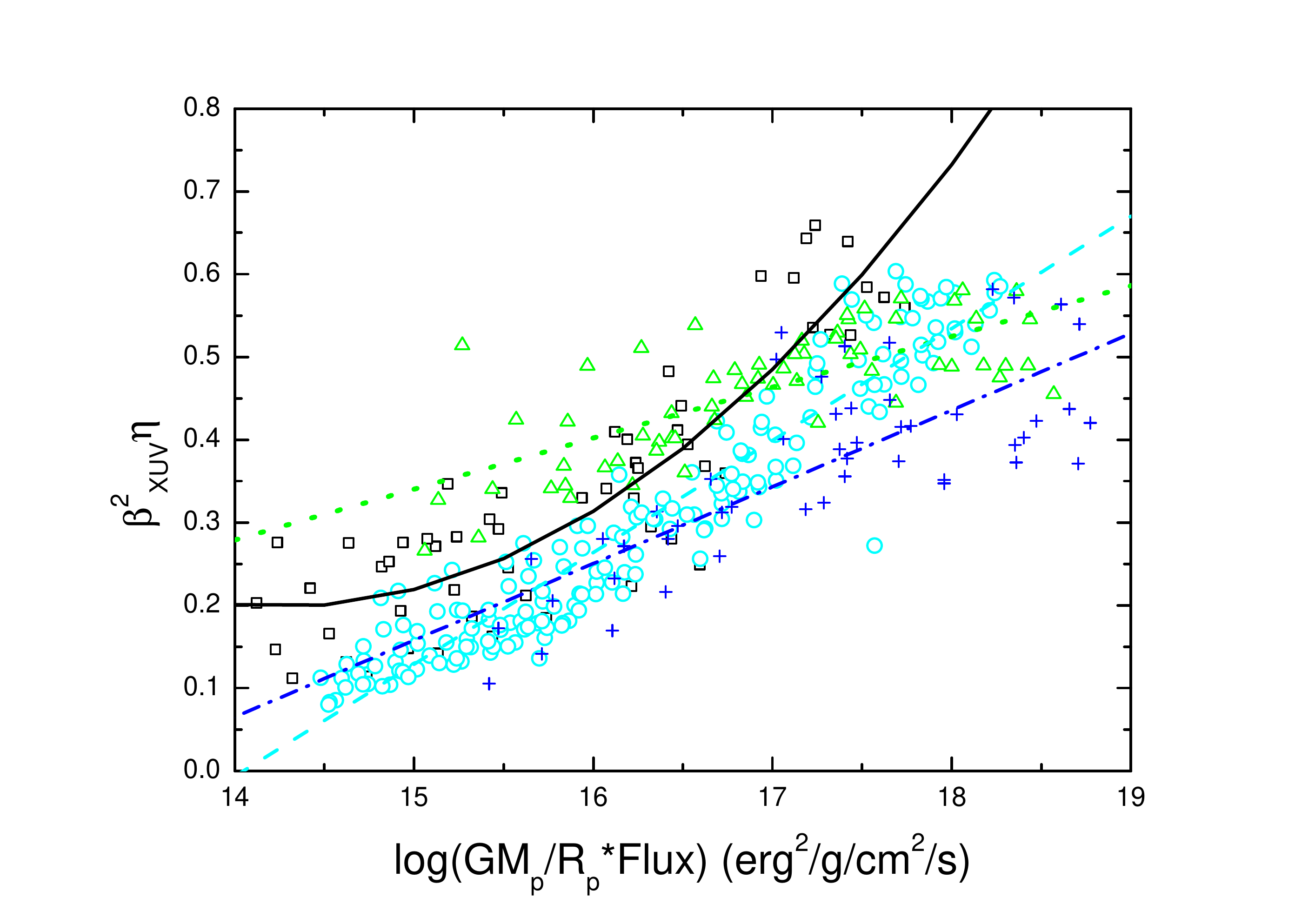}
\caption{The fit of $\beta_{xuv}^{2}$ $\eta$ for planets with the gravitational potentials smaller than 1.5$\times$10$^{13}$ erg g$^{-1}$. The x-axis is product of stellar irradiation and planetary gravitational potential, and y-axis is the $\beta_{xuv}^{2}$ $\eta$. Black square: the planets with the gravitational potential smaller than 1.5$\times$10$^{12}$ erg g$^{-1}$. Cyan circle: the gravitational potential are between 1.5$\times$10$^{12}$ erg g$^{-1}$ and 5$\times$10$^{12}$ erg g$^{-1}$. Green triangle: the gravitational potential are between 5$\times$10$^{12}$ erg g$^{-1}$ and 10$^{13}$ erg g$^{-1}$. Blue cross: the gravitational potential are between 1$\times$10$^{13}$ erg g$^{-1}$ and 1.5$\times$10$^{13}$ erg g$^{-1}$. The colored lines are the fit to the corresponding $\beta_{xuv}^{2}$ $\eta$.}
\end{figure}

When the GM$_{p}$/R$_{p}$*F$_{XUV}$ is given, there is a trend that the values of $\beta_{xuv}^{2}$ $\eta$ decrease with the increase of gravitational potential except the third group. We found that the distributions of $\beta_{xuv}^{2}$ $\eta$ of the third group are higher than those of adjacent groups. The gravitational potentials of these planets (5$\times$10$^{12}$ erg g$^{-1}$ - 10$^{13}$ erg g$^{-1}$) are in an intermediate range, but the slope of this line is the smallest. To explain the issue, we checked the dependence of the XUV absorption radius $\beta_{xuv}$ and heating efficiency $\eta$ on the gravitational potential (see Figure. 4 and Figure. 5). For planets with gravitational potentials smaller than 1.5$\times$10$^{12}$ erg g$^{-1}$, $\beta_{xuv}$ is in the range of 1.1-1.7. Their values of $\eta$ vary from 0.05 to 0.45. Thus, $\beta_{xuv}^{2}$ $\eta$ covers a broad range. For planets with gravitational potentials higher than 1.5$\times$10$^{12}$ erg g$^{-1}$, $\beta_{xuv}$ is in a small range (1.03-1.3). However, the heating efficiencies are different for different planets. For the second and fourth group, the distributions of heating efficiency are similar with those of the first group. Due to the smaller $\beta_{xuv}$, their $\beta_{xuv}^{2}$ $\eta$ are smaller than those of the first group. Furthermore, $\beta_{xuv}$ will decrease with the increase of gravitational potentials so that the values of $\beta_{xuv}^{2}$ $\eta$ of the fourth group are smaller than those of the second group. However, the planets in the third group (most planets in the third group is Jupiter-like planets) are close to the transitional regions of heating efficiency (see the right panel of Figure. 4) where the heating efficiencies maintain higher values (0.3-0.45). Compared with the second and fourth group, their XUV absorption radii are similar. Thus, the high heating efficiency of the third group caused the high values of $\beta_{xuv}^{2}$ $\eta$ and a smaller slope of the third line.

\subsection{\textbf{The influence of the outer boundary on absorption depth}}

In the simulations, the atmospheric outer boundaries of planets are set to equal the stellar radii. However, the real boundaries can not be determined well because the atmosphere of planet can be constrained to a few or tens planetary radii by the stellar wind. For planets with the size of Jupiter, the total pressure of the stellar wind can be balanced at about a few planetary radii by the ram pressure and the thermal pressure of planet\citep{2009APJ...693..23}. Thus, the outer boundaries can roughly be denoted by the radius of host star. However, the calculating outer boundaries of the Earth-like planets could be far larger than their real boundary or magnetosphere due to the decrease of planetary pressure with the increase of radius. Here we inspected the dependence of the absorption of Ly$\alpha$ on atmospheric outer boundaries by some planets of our sample. The effect of outer boundary on the Ly$\alpha$ absorption is shown in Figure. 11. In the left panel of Figure. 11, we investigated the cases of Jupiter-like and Saturn-like planets. The right panel showed the dependence of the absorption on outer boundary for Neptune-like and Earth-like planet.

As we can see, both the absorptions increase with the increasing atmospheric boundaries. For most Jupiter-like planets, the increase is prominent with the increase of outer boundary so that the maximum absorption can attain 30\%-50\%. For the planets with the size of Neptune and Earth, the absorptions produced within 10 R$_{p}$ are smaller than 10\% (even the absorptions of some Earth-like planet are negligible). With the increase of the radii, the absorption of some planets can attain 15\%. If the real boundary or magnetosphere of Earth-like planet are smaller than the radii of their host stars, this hints that the absorptions of Earth-like planets in Figure. 7 can be overestimated. In addition, the absorptions of some Earth-like planets are almost insensitive to the outer boundary. As shown in the right panel of Figure. 11, the increase of a factor of a few in the outer boundary only results in a few percent raise in Ly$\alpha$ absorption. Compared to the case of Jupiter-like planets, it is obvious that detecting the absorption signals of Ly$\alpha$ in Earth-like planets is not easy.
\begin{figure}
\includegraphics[scale=0.45,trim=10 320 10 0]{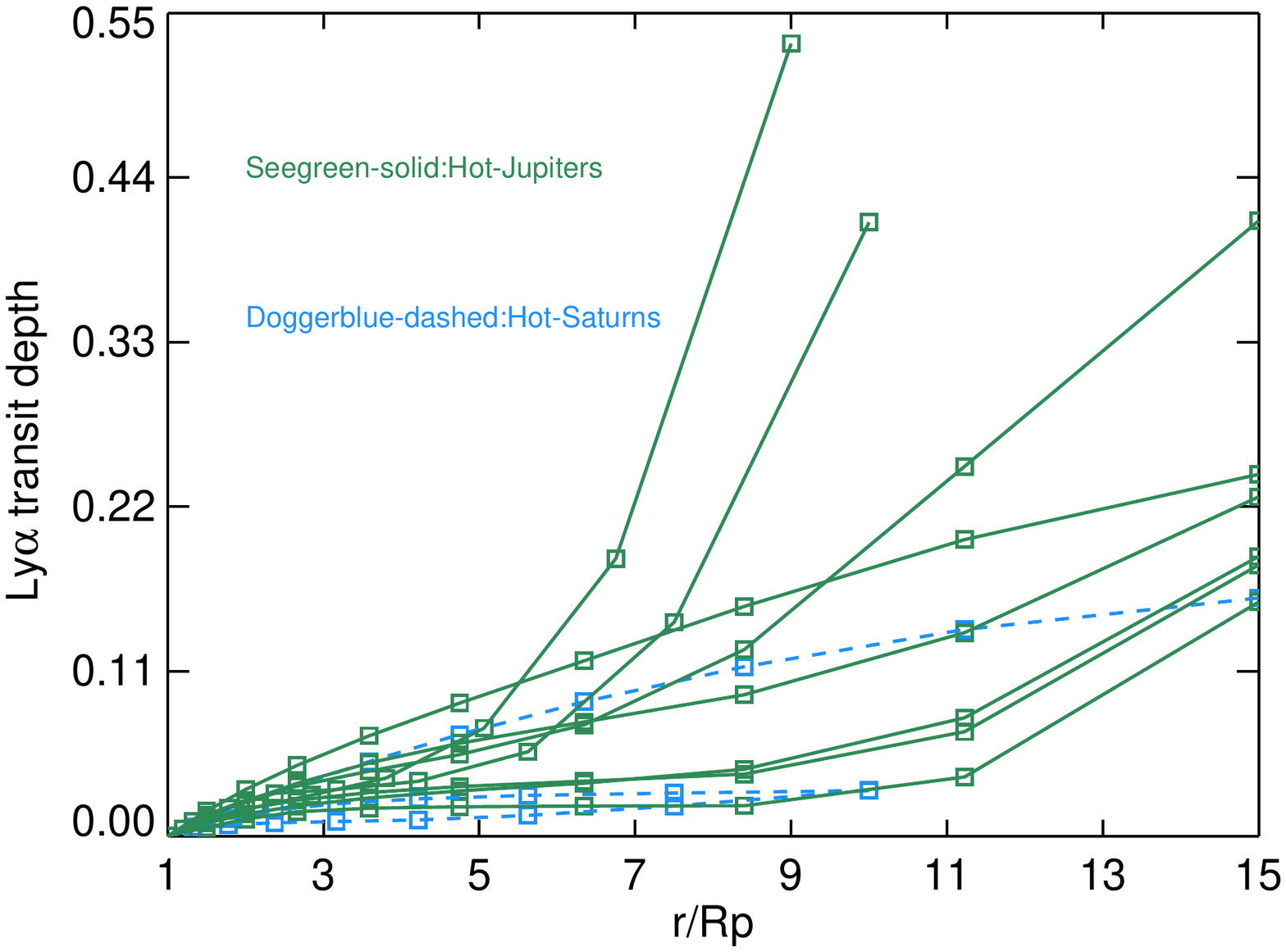}
\includegraphics[scale=0.45,trim=10 320 1130 0]{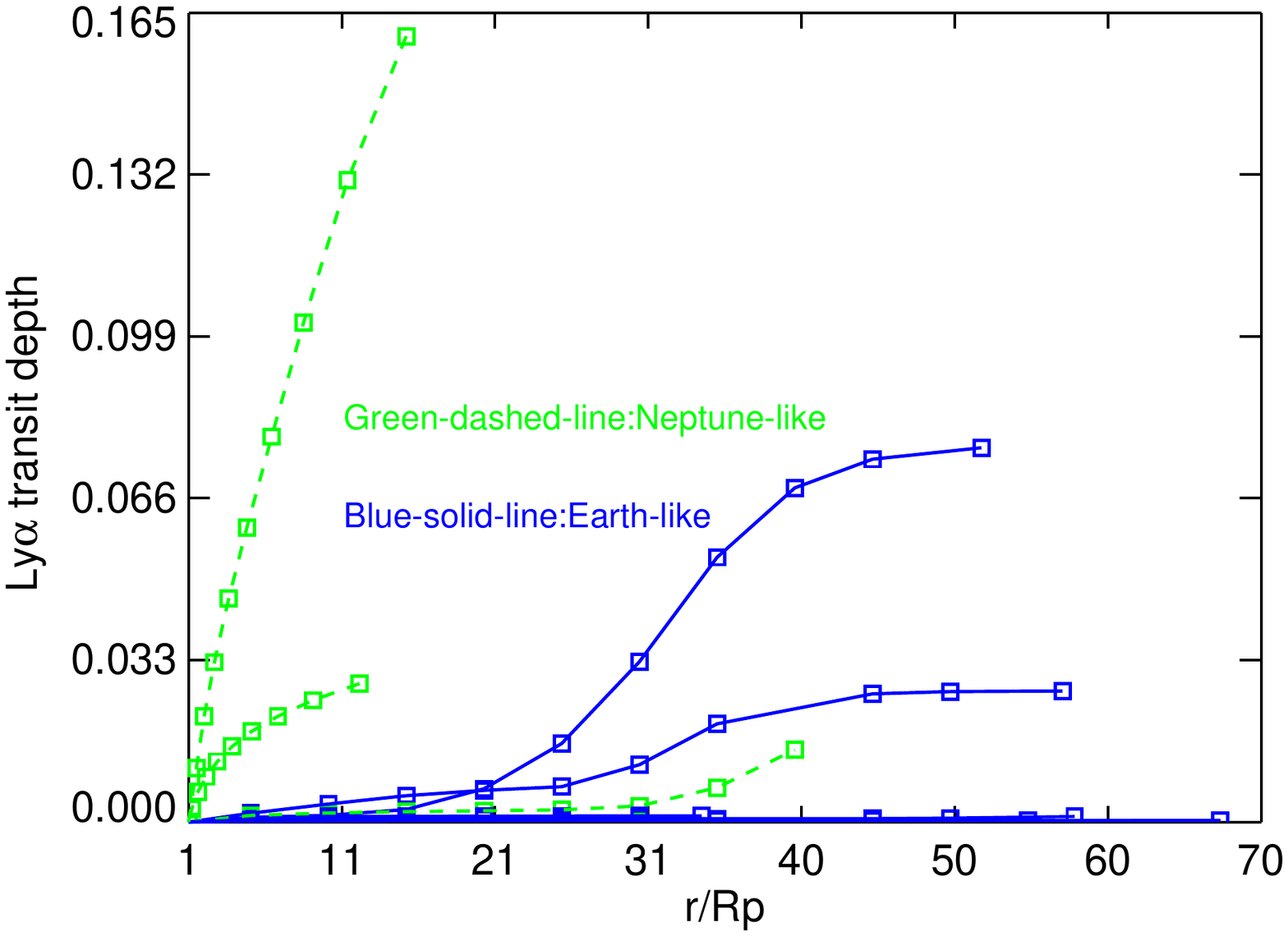}
\caption{L$\alpha$ transit depths in line wing([1215.06 $\rm\AA$, 1215.47 $\rm\AA$] $\cup$ [1215.87 $\rm\AA$, 1216.28 $\rm\AA$]) versus outer boundary.  The left panel denotes the average transit depths for Jupiter-like and Saturn-like planets. The right panel shows the average transit depths for Neptune-like and Earth-like planets. \label{boundarytransit}}
\end{figure}

\subsection{\textbf{The limitation of this work}}\label{subsec:futurework}

In our work, the mechanism of atmospheric escape is thermal escape due to intense stellar XUV radiation. We account for the photochemical interactions of planetary particles. However, other aspects such as the influence of interplay between stellar wind and planetary wind and the impact of planetary magnetism are not taken into consideration. Moreover, this work is based on 1D simulations of hydrodynamic atmospheric escape. For tidally-locking planets, the results may lead to some deviation. Therefore, in our future study, we are going to take these factors into account. Furthermore, in this paper, we only focus on the absorption of Ly$\alpha$ by the atmosphere of planet. It is the first step towards predicting the observable signals because we do not include the influence of charge-exchange and the extinction of ISM. In addition, an accurate XUV spectra is also needed if the observable signals of some special targets need to be known. Finally, the excess absorption depth of stellar Ly$\alpha$ by planetary atmosphere we predicted here can provide some clues for future observations.

\section{Summary} \label{sec:summ}
In this paper, we investigated about 450 transit systems. We obtained the atmospheric structures of our selected planets based on our 1D hydrodynamic atmospheric escape simulations \citep{2011ApJ...733..98,2013ApJ...766..102,2016ApJ...818..107,2018CHA&A...42..81} and simulate the absorption of stellar Ly$\alpha$ by these planets' atmosphere. Based on the simulations, we found that the mass loss rates are dependent of the mean density and the XUV irradiation. Our results suggest the energy-limited assumption reflects the essential physics of the hydrodynamic escape of the atmosphere. However, the energy-limited equation can overestimate the mass loss rates due to the neglect of the kinetic and thermal energy of the escaping atmosphere. We found that the overestimation is prominent for planets with smaller sizes. For Jupiter-like planets, the deviation of the mass loss rates are lower due to their large gravitational potential. By correcting the kinetic and thermal energy and using the heating efficiency and the absorption radius of XUV irradiation of our hydrodynamic model, the results of our hydrodynamic mass loss rates are consistent with those of revised energy-limited equation. We calculated the heating efficiency and XUV absorption radius for each planet. The heating efficiency are almost proportional to the logarithm of the product of the XUV flux and the gravitational potential(i.e., log(F$_{xuv}GM_p/R_p$)). The R$_{XUV}$ tended to be higher when the planetary radii and masses are smaller.  Finally, in order to use the energy-limited equation easily we fitted the $\beta^{2}_{XUV}\eta$ by using our results.

In addition, we obtained some statistical properties about the distribution of the Ly$\alpha$ absorption depth. We found that the absorption depth would be larger if the planetary mean densities are lower and the integrated XUV Flux are higher. This means that the planets with lower mean densities and subjected to more intense Fxuv are likely to increase their excess absorption depth.
Moreover, different absorption levels could be approximately divided by different mass loss rates. The higher the mass loss rates, the higher the absorption depth. The obvious absorptions will appear when mass loss rates are higher than 10$^{11}$ g/s. For the case of lower mass loss rates, the absorption  decreases to a low level. Finally, the strong absorption levels appear in the planets with large size, which can be attribute to the higher mass loss rates of those planets.

We thank the referee for their comments and
suggestions, which helped to improve the quality of this work. For
the author, this work is supported by National Natural Science Foundation of China (Nos.11273054; Nos.11333006) and by the project `` Technology of Space Telescope Detecting Exoplanet and Life " from National Defense Science and Engineering Bureau civil spaceflight advanced research project (D030201). This work has made use of the MUSCLES Treasury Survey High-Level Science Products; doi:10.17909/T9DG6F.

\end{document}